\documentclass{JHEP3}

 \usepackage[english]{babel}
 \usepackage{cite}
\usepackage{subfigure}
\usepackage{epsf}
 \usepackage{amssymb}
 \usepackage{amsmath}
 \usepackage{amsfonts}
 \usepackage{psfrag,epsfig,graphicx,graphics}

\def\nn{\nonumber}
\def\bei{\begin{itemize}}
\def\ei{\end{itemize}}
\def\beqa{\begin{eqnarray}}
\def\eqa{\end{eqnarray}}
\def\bea{\begin{eqnarray}}
\def\eea{\end{eqnarray}}
\def\beas{\begin{eqnarray*}}
\def\eeas{\end{eqnarray*}}
\def\beqas{\begin{eqnarray*}}
\def\eqas{\end{eqnarray*}}

\def\beq{\begin{equation}} 
\def\be{\begin{equation}}

\def\ee{\end{equation}}
\def\eq{\end{equation}}
\def\eeq{\end{equation}}

\def\beqd{\begin{displaymath}}
\def\eeqd{\end{displaymath}}
\def\eqd{\end{displaymath}}

\def\beeq{\begin{eqnarray}} \def\eeeq{\end{eqnarray}}

\def\bef{\begin{frame}}

\def\slashchar#1{\setbox0=\hbox{$#1$}
   \dimen0=\wd0
   \setbox1=\hbox{/} \dimen1=\wd1
   \ifdim\dimen0>\dimen1
      \rlap{\hbox to \dimen0{\hfil/\hfil}}
      #1
   \else
      \rlap{\hbox to \dimen1{\hfil$#1$\hfil}}
      /
   \fi}

\usepackage{graphicx}
\usepackage{dcolumn}
\usepackage{bm}
\usepackage{epsfig}

\newcommand{\dhd}{{\textstyle d}
\lower.03ex\hbox{\kern-0.38em$^{\scriptstyle-}$}\kern-0.05em{}}
\newcommand{\dbar}{{\textstyle \delta}
\lower.03ex\hbox{\kern-0.38em$^{\scriptstyle-}$}\kern-0.05em{}}

\newcommand{\qb}{{\bar q}}
\newcommand{\sla}{\slash \!\!\!}
\newcommand{\eg}{\underline{e}}

\newcommand{\eb}{{\underline e}}

\newcommand{\Db}{{\underline \Delta}}
\newcommand{\kb}{{\underline k}}

\newcommand{\yb}{{\bar y}}

\newcommand{\rb}{{\underline r}} 
\title{Saturation effects in exclusive $\rho_{T,L}$ meson electroproduction}
\author{A. Besse\\
LPT, Universit{\'e} Paris-Sud, CNRS, 91405, Orsay, France \\
Email: \email{besse@th.u-psud.fr}}
\author{ L. Szymanowski\\
National Center for Nuclear Research (NCBJ), Warsaw, Poland\\
Email: \email{Lech.Szymanowski@fuw.edu.pl}}
\author{S. Wallon\\
LPT, Universit{\'e} Paris-Sud, CNRS, 91405, Orsay, France {\em \&} \\
UPMC Univ. Paris 06, facult\'e de physique, 4 place Jussieu, 75252 Paris Cedex 05, France
Email: \email{wallon@th.u-psud.fr}}

\abstract{
We use recent results for the $\gamma^*_{L}\to\rho_L$ and $\gamma^*_T\to\rho_T$ impact factors, computed 
in the impact parameter representation within the collinear factorization scheme, to get predictions for the polarized cross-sections $\sigma_{T}$ and $\sigma_L$ of the diffractive leptoproduction of the $\rho$ meson at high energy. 
In this approach the helicity amplitude is a convolution of the scattering 
amplitude of a color dipole with a target, together with the virtual gamma 
wave 
function and with the first moments of the $\rho$ meson wave function
(in the transverse momentum space), given by
the distribution amplitudes up to twist~3 for the $\gamma^*_T\to\rho_T$ impact factor and up to twist~2 for the $\gamma^*_L\to\rho_L$ impact factor. 
 Combining these results 
 with recent dipole models fitted to DIS data, which include saturation effects, we show that the predictions are in good agreement with HERA data for photon virtuality ($Q^2$) larger than typically 5\;GeV$^2$, without free parameters and with a weak dependence on the choice of the factorization scale, i.e. the shape of the DAs, for both longitudinally and transversely polarized $\rho$ meson. For lower values of $Q^2$, the inclusion of saturation effects is not enough to provide a good description of HERA data. We believe that it is a signal of a need for higher twist contributions in the $\rho$ meson DAs. We also analyze the radial distributions of dipoles between the initial $\gamma^*$ and the final $\rho$ meson states.}

\date{\today}
\begin{document}

\pagestyle{empty}
\newpage

\mbox{}

\pagestyle{plain}

\setcounter{page}{1}


\section{Introduction}
\label{sec:intro}
In the high energy limit, exclusive processes and more particularly the diffractive leptoproduction of vector mesons, provide a nice probe to study the hadronic properties. From the experimental side, data have been extracted
in a wide range of center-of-mass energies, from a few GeV at JLab to hundreds of GeV at
the HERA collider. The kinematical range which is at the heart of the present paper is the large energy in the center-of-mass of $\gamma^* p$, denoted $W$, for which 
the HERA collaborations H1 and ZEUS measured $\rho$-meson electroproduction starting from the early period of HERA activity \cite{Breitweg:1998nh,  Breitweg:1999fm} till now, with an increasing precision leading
 to a complete analysis  \cite{Chekanov:2007zr, Aaron:2009xp} of spin density matrix elements, polarized  and total cross-sections describing the hard exclusive productions of the $\rho$ and the $\phi$ vector mesons $V$ in the process
\begin{equation}
\label{process}
\gamma^{*}(\lambda_{\gamma}) \, p \rightarrow V(\lambda_{V}) \, p \,.
\end{equation}
These matrix elements and polarized cross-sections
can be expressed in terms of  helicity amplitudes $T_{\lambda_{V}\lambda_{\gamma}}$ ($\lambda_{\gamma}$, $\lambda_{V}$ : polarizations of the virtual photon and of the vector meson).

The ZEUS collaboration \cite{Chekanov:2007zr} has provided data for different photon virtualities $Q^2$, i.e. for
 $2 < Q^2 < 160$ GeV$^2, \, 32 < W < 180$ GeV $(\;|t| < 1$ GeV$^2),$ while the H1 collaboration \cite{Aaron:2009xp} has analyzed data
in the range
$2.5 < Q^2 < 60$ GeV$^2, \, 35 <  W < 180$ GeV  $\;(|t| < 3$ GeV$^2)\,.$  
 The high virtuality of the exchanged photon 
 allows the factorization of the amplitude into a hard subprocess described within the perturbative QCD approach and suitably defined hadronic objects, such as the dipole-nucleon scattering amplitude or the vector meson wave functions and distribution amplitudes (DAs) \cite{Donnachie:1987pu,Nemchik:1996pp,Munier:2001nr}. HERA data are thus interesting observables to test the properties of these non-perturbative objects such as the saturation dynamics of the nucleon or the transverse momentum dependence of the vector meson wave functions.

On the theoretical side, three main approaches have been developed.  The first two, a $k_T$-factorization approach and a dipole approach,  are applicable at high energy, $W\gg Q \gg \Lambda_{QCD}$. They are both related to a
Regge inspired $k_T$-factorization scheme \cite{Cheng:1970ef, FL, Gribov:1970ik, Catani:1990xk, Catani:1990eg, Collins:1991ty, Levin:1991ry}, which basically writes the scattering amplitude in terms of two impact
factors: one, in our case,  for the $\gamma^* - \rho$ transition and the other one for the nucleon to nucleon transition,
with, at leading order, a two "Reggeized" gluon exchange in the $t$-channel. The Balitsky-Fadin-Kuraev-Lipatov (BFKL)  evolution, known at leading order (LLx) \cite{Fadin:1975cb, Kuraev:1976ge, Kuraev:1977fs, Balitsky:1978ic} and next-to-leading (NLLx) order \cite{Fadin:1996tb, Camici:1997ij, Ciafaloni:1998gs, Fadin:1998py}, can then be applied
to account for a specific large energy QCD resummation.
The dipole approach is  based on  the formulation of similar ideas althought not in $k_T$ but in transverse coordinate space \cite{Mueller:1989st, Nikolaev:1990ja}; this scheme is especially suitable to account for nonlinear evolution and gluon saturation effects.
The third approach, valid down to $W\sim Q$, was initiated in \cite{Brodsky:1994kf} and \cite{Frankfurt:1995jw}.
It is based on
the collinear QCD factorization scheme \cite{Collins:1996fb, Radyushkin:1997ki}; the amplitude is given as  a convolution of quark or gluon generalized parton
distributions (GPDs) in the nucleon, the $\rho$-meson DA, and a perturbatively calculable
hard scattering amplitude. GPD evolution equations resum the collinear quark and gluon effects. The DAs are also subject to specific QCD evolution equations \cite{Farrar:1979aw,  Lepage:1979zb, Efremov:1979qk}.

Though the collinear factorization approach allows us to calculate perturbative corrections to the leading twist longitudinal amplitude (see \cite{Ivanov:2004zv} for NLO), when dealing with transversely polarized vector mesons related to higher twist 
contributions, one faces end-point singularity problems. Consequently, this does not allow us to study polarization effects in diffractive $\rho$-meson electroproduction in a model-independent way within the collinear factorization approach. An improved collinear approximation scheme \cite{Li:1992nu} has been proposed, 
 which allows us to overcome end-point singularity problems, and which has been applied to $\rho$-electroproduction \cite{Vanderhaeghen:1999xj, Goloskokov:2005sd, Goloskokov:2006hr, Goloskokov:2007nt}.

In this study, we consider polarization effects for the reaction (\ref{process}) in the high energy region, $s=W^2\gg Q^2\gg \Lambda^2_{QCD}$,
working within the $k_T$-factorization approach, where 
the helicity amplitudes can be expressed\footnote{We use underlined letters for Euclidean two-dimensional transverse vectors.} as the convolution of the $\gamma^*\to\rho$ impact factor $\Phi^{\gamma^{*}(\lambda_{\gamma}) \rightarrow \rho(\lambda_{\rho})}(\underline{k}^2,Q^2)$ with the unintegrated gluon density ${\cal F}(x,\underline{k}^2 )$ density, which at the Born order is simply related to the nucleon impact factor $\Phi^{N\to N}(\underline{k}^2,Q^2)$, where $\underline{k}$ is the transverse momentum of the $t-$channel exchanged gluons. 
 For the transverse amplitude the end-point singularities are naturally regularized by the transverse
momenta of the $t-$channel gluons \cite{Ivanov:1998gk, Anikin:2009hk, Anikin:2009bf}. At large photon virtuality, the $\gamma^* - \rho$ impact factors can be calculated in a model-independent way using QCD twist expansion in the region $\underline{k}^2\gg \Lambda_{QCD}^2$. Such a calculation  involves the $\rho$-meson DAs as nonperturbative inputs. 
The calculation of the impact factors for $\Phi^{\gamma^{*}_L \rightarrow \rho_L}$, $\Phi^{\gamma^{*}_T \rightarrow \rho_L}$
is standard at the twist-2 level \cite{Ginzburg:1985tp}, the next term of the expansion being of twist~4, while $\Phi^{\gamma^{*}_T \rightarrow \rho_T}$ was only recently computed \cite{Anikin:2009hk,Anikin:2009bf} (for the forward case $t = t_{\rm min}$),  up to twist-3, including two- and  three-parton correlators, which contribute here on an equal footing.

In a previous study~\cite{Anikin:2011sa}, we used the results  \cite{Ginzburg:1985tp, Anikin:2009bf} for the $\Phi^{\gamma^*_L \rightarrow \rho_L}$ and  $\Phi^{\gamma^*_T \rightarrow \rho_T}$ impact factors and a phenomenological model \cite{Gunion:1976iy} for the proton impact factor. It was pointed out that the region $\underline{k}^2\gg \Lambda_{QCD}^2$ gives the dominant contribution to the helicity amplitudes 
 while the soft gluon ($\underline{k}^2<1\;$GeV$^2$) contribution cannot be neglected. The soft gluon contribution to the amplitudes involves the interaction of large size color dipole configuration $|\rb|$ ($|\rb|\equiv 1/|\kb|$) in the fluctuations of the probe and saturation effects could then play an important role.

Following this idea, we have shown in ref.~\cite{Besse:2012ia} that the helicity amplitudes, expressed in impact parameter space and then computed in the collinear factorization scheme, factorize into the dipole cross-section and the wave functions of the virtual photon combined with the first moments of the $\rho$ meson wave functions parameterized by the DAs, given by the twist expansion up to twist~3 for the production of a $\rho_T$ and to twist~2 when producing a $\rho_L$. Note that in \cite{Ivanov:1998gk} a very similar approach was followed, where the DAs were replaced by the Taylor expansion of models of the $\rho$ meson wave function at small $q\qb$ pair transverse size. 
The results of ref.~\cite{Besse:2012ia} link the $k_T-$factorization approach, in particular the results of \cite{Anikin:2009bf}, with 
 the calculations performed in refs.~\cite{Nemchik:1996cw, Forshaw:2003ki, Kowalski:2006hc, Forshaw:2010py, Forshaw:2011yj} within the dipole approach. The main difference between our present approach and the one of refs.~\cite{Nemchik:1996cw, Forshaw:2003ki, Kowalski:2006hc, Forshaw:2010py, Forshaw:2011yj} is that instead of using the  
light-cone wave functions $\phi(z,\rb)$ which in practice need to be modeled, the amplitude of our approach involves the DAs which parameterize the first moments of the wave functions. 

Our main point here is that one can assume the dominant physical mechanism for production of both longitudinal and transversely polarized mesons to be the scattering of small transverse-size quark-antiquark and
quark-antiquark-gluon colorless states on the target. This allows to calculate corresponding helicity amplitudes 
in a model-independent way, using the natural light-cone QCD language --  twist-2 and twist-3 DAs.

This paper is organized as follows. In  section~\ref{sec:IF_rep}, we introduce the impact factor representation of the helicity amplitudes as well as the kinematics of the process. In section~\ref{sec:IF}, we first recall some results for the $\gamma^*_L\to\rho_L$ and $\gamma^*_T\to\rho_T$ impact factors computed in momentum space respectively up to twist~2~\cite{Ginzburg:1985tp} and twist~3~\cite{Anikin:2009bf} accuracies using the collinear approximation to parameterize the soft part associated to the production of the $\rho$ meson by distribution amplitudes (DAs), calculated in ref.~\cite{Ball:1998sk,Ball:2007zt}. We recall then the expression of these impact factors in the impact parameter space according to the results of ref.~\cite{Besse:2012ia} allowing to decouple the dipole-nucleon scattering amplitude from the amplitude of production of dipoles in the initial ($\gamma^*(\lambda_{\gamma})$) and final ($\rho(\lambda_{\rho})$) states. We terminate section~\ref{sec:IF} by expressing helicity amplitudes and polarized cross-sections in terms of the dipole cross-section. 
 In section~\ref{sec:Dipole models}, we give a brief review of the main properties of the models for the dipole cross-section \cite{GolecBiernat:1998js,Albacete:2007yr,Albacete:2007sm} or the proton impact factor \cite{Gunion:1976iy} that we use in our study. 
We compare our predictions with the data of HERA\cite{Chekanov:2007zr,Aaron:2009xp} in section~\ref{sec:HERADATA} and we obtain a good agreement. In this context we discuss the role of higher twist corrections for small $Q^2$ values. Finally, we analyse the radial distribution of dipole intermediate states involved between the virtual photon and the $\rho$ meson, and discuss the role of the saturation models on the specific example of the Golec-Biernat and W\"usthoff saturation model \cite{GolecBiernat:1998js}. We also compare our radial distribution with the overlap of the $\gamma^*$ and $\rho$ meson wave functions obtained in the approach of the dipole models \cite{Forshaw:2003ki,Kowalski:2003hm,Kowalski:2006hc}, where the $\rho$ meson wave function is modeled and the parameters are fitted to HERA data. 

\section{Helicity amplitudes of the hard $\rho$ meson leptoproduction in the impact factor representation}
\label{sec:IF_rep}
\begin{figure}[h]
\psfrag{Kap}[cc][cc]{\hspace{.3cm}\raisebox{-.7cm}{$\kappa$}\rotatebox{-60}{$\underbrace{\rule{0.8cm}{0pt}}$}}
\psfrag{K}[cc][cc]{$k$}
\psfrag{KR}[cc][cc]{$k-\Delta$}
\psfrag{Rho}[cc][cc]{$\rho (p_{\rho})$}
\psfrag{Gams}[cc][cc]{$\gamma^*(q)$}
\psfrag{N}[cc][cc]{$N$}
\psfrag{Np}[cc][cc]{$N$}
\psfrag{PHIGAM}[cc][cc]{$\Phi^{\gamma^*\to\rho}$}
\psfrag{PHINN}[cc][cc]{$\Phi^{N\to N}$}
\centerline{\raisebox{0cm}{\epsfig{file=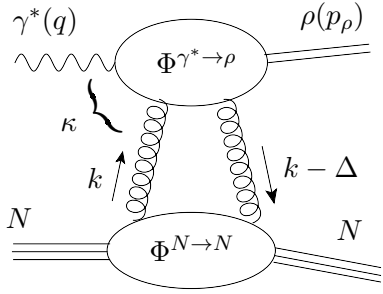,width=5cm,clip=}}}
\caption{Impact factor representation of the $\gamma^* \, N  \to  \rho \, N$ scattering amplitude.}
\label{Fig:impact_fact}
\end{figure}

In the impact factor representation at the Born order, the amplitude of the exclusive process $\gamma^{*}(\lambda_{\gamma}) \, N \rightarrow \rho (\lambda_{\rho}) \, N$ reads
\begin{equation}
\label{defImpactRep}
 T_{\lambda_{\rho}\lambda_{\gamma}}(\Db;Q , M) = is\int \frac{d^2\underline{k}}{(2\pi)^2}\frac{1}{\underline{k}^2(\underline{k}-\Db)^2} \Phi^{N \rightarrow N} (\underline{k},\Db;M^2)\Phi^{\gamma^{*}(\lambda_{\gamma}) \rightarrow \rho(\lambda_{\rho})}(\underline{k},\Db;Q^2)\,,
 \end{equation}
as illustrated in figure~\ref{Fig:impact_fact}.
The $\gamma^{*}(\lambda_{\gamma}) \rightarrow \rho(\lambda_{\rho})$ impact factor
$ \Phi^{\gamma^{*}(\lambda_{\gamma}) \rightarrow \rho(\lambda_{\rho})}$ is defined through the discontinuity of the  $S$ matrix element for $\gamma^{*}(\lambda_{\gamma};q) g(k) \rightarrow g(k-\Delta) \rho(\lambda_{\rho};p_{\rho})$
as
\beq
\label{ImpactDisc}
\Phi^{\gamma^{*}(\lambda_{\gamma}) \rightarrow \rho(\lambda_{\rho})}=\frac{1}{2 s}\int \frac{d \kappa}{2 \pi} \, {\rm Disc}_\kappa  \left( S_{\mu \nu}^{\gamma^* g \to \rho g} \, p_2^\mu \, p_2^\nu \frac{2}{s} \right)\,,
\eq
where $
\kappa =(k+q)^2$\,. In eqs.~(\ref{defImpactRep}) and (\ref{ImpactDisc}) the momenta $q$ and $p_\rho$ are parameterized via Sudakov decompositions, in terms of two
 lightlike vectors $p_1$ and $p_2$ such that $2\, p_1.p_2=s$, as
\begin{equation}
q=p_1-\frac{Q^2}{s}p_2 \quad {\rm and } \quad p_{\rho}=p_1+\frac{m_{\rho}^2-t+t_{\rm min}}{s} p_2 + \Delta_\perp\,,
\end{equation}
where $Q^2=-q^2 >> \Lambda_{QCD}^2$ is the virtuality of the photon being the large scale which justifies the use of perturbation theory, and $m_{\rho}$ is the mass of the $\rho$ meson. 
Here $-t_{\rm min}$ denotes the minimal value of $-t\,.$
The nucleon impact factor
$ \Phi^{N \rightarrow N}$ in eq.~(\ref{defImpactRep}) cannot be computed within perturbation theory, and is related\footnote{Normalization of the impact factors differs from \cite{Forshaw:1997dc} by a factor $2\pi$, $\Phi^{\text{\cite{Forshaw:1997dc}}}=2\pi \Phi^{Here}$. For clarity, we have restored in eq.~(\ref{conv-phi-F1}) 
the colored indiced carried by the impact factor $\Phi_{ab}^{\gamma^{*}_{\,\lambda_{\gamma}}\rightarrow \rho_{\lambda_{\rho}}}$.} at Born order to the unintegrated gluon density $\mathcal{F}(x,\kb)$. In the forward limit $\Delta_{\perp}=0$, the helicity amplitudes read,
\beq
\frac{T_{\lambda_{\rho},\lambda_{\gamma}}}{s}
=\frac{\delta^{ab}}2 \int \frac{d^2 \kb}{\kb^4} \Phi_{ab}^{\gamma^{*}_{\lambda_{\gamma}} \rightarrow \rho_{\lambda_{\rho}}}(\kb,Q,\mu_F^2) \mathcal{F}(x,\kb)\,.
\label{conv-phi-F1}
\eq
The impact factors $\Phi^{\gamma^{*}(\lambda_{\gamma}) \rightarrow \rho(\lambda_{\rho})}(\underline{k},Q,\mu_F^2)$ and the nucleon impact factor vanish at $\underline{k}\to 0$ or $\kb\to\underline{\Delta}$, 
which guarantees the convergence of the integral in eq.~(\ref{conv-phi-F1}) on the lower limit\footnote{This property of the impact factors is universal in the case of the scattering of colorless objects and is related to gauge invariance \cite{Frolov:1970ij,Fadin:1999qc}.}

\begin{figure}[h]
\begin{tabular}{cccc}
\includegraphics[width=5.8cm]{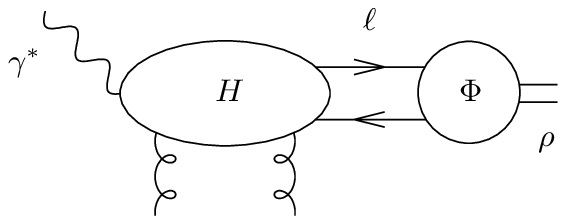}&\hspace{.2cm}
\raisebox{1.2cm}{+}&\hspace{.3cm}\includegraphics[width=5.8cm]{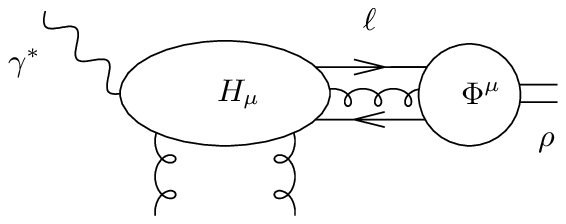}
&\hspace{.2cm}\raisebox{1.2cm}{$+ \cdots$}
\end{tabular}
\caption{Two- and three-parton correlators attached to a hard scattering amplitude in the  specific case of the $\gamma^* \to \rho$ impact factor,
where vertical lines are hard $t-$ channel gluons in the color
singlet state.}
\label{Fig:NonFactorized}
\end{figure}

The computation of the  $\gamma^* \to \rho$ impact factor is performed within collinear factorization of QCD.
The dominant contribution corresponds to the $\gamma^*_L \to \rho_L$ transition (twist~2), while the other transitions  are power suppressed.
The $\gamma^*_L \to \rho_L$ and  $\gamma^*_T \to \rho_L$ impact factors were computed a long time ago \cite{Ginzburg:1985tp}, while a consistent treatment of the twist-3
$\gamma^*_T \to \rho_T$ impact factor has been performed only recently in ref.~\cite{Anikin:2009bf}.
It is based on the light-cone collinear factorization (LCCF) beyond the leading twist, applied to the amplitudes $\gamma^{*}(\lambda_{\gamma})g(k) \rightarrow g(k-\Delta)\rho(\lambda_{\rho})$, symbolically illustrated in figure~\ref{Fig:NonFactorized}.
Each of these scattering amplitudes is the sum of the convolutions of a hard part (denoted by $H$ and $H_\mu$ for two- and three-parton contributions, respectively) that corresponds to the transition of the virtual photon into the constituents of the $\rho$ meson and their interactions with off-shell gluons of the $t$ channel, and a soft part (denoted by $\Phi$ and $\Phi^\mu$) describing the hadronization of the constituent partons into the $\rho$ meson and parameterized up to a given twist by DAs.


\section{Helicity amplitudes and polarized cross-sections}
\label{sec:IF}
\subsection{Impact factors  $\gamma^{*}_{L,T} \rightarrow \rho_{L,T}$}
\label{subsec:IFM}
In the Sudakov basis,
the longitudinal and
 transverse polarizations   of the photon are\footnote{In ref.~\cite{Anikin:2009bf} we took  $\epsilon^{\pm}=\mp\frac{i}{\sqrt{2}}\left(0, 1,\pm i, 0\right)$, which we change here for consistency with the usual experimental conventions \cite{:2010dh}.}
\beq
\label{def_polL_polT}
e_{\gamma L}^{\mu}=\frac{1}{Q} (p_1^{\mu}+\frac{Q^2}{s}p_2^{\mu})\,,
\qquad
 \epsilon^{\pm}=\frac{1}{\sqrt{2}}\left(0, \mp1,- i, 0\right)\,.
\eq
For $t=t_{\rm min}$ the same parametrization will be used for the $\rho$-meson polarization with obvious substitutions $Q^2 \to -m_{\rho}^2$ and $Q \to m_{\rho}$. 

Let us introduce $p$ and $n$, two light-cone vectors such that $p_{\rho}\approx p$ at twist 3 and $p \cdot n = 1$. The polarization of the out-going $\rho$ meson is denoted by $e^*$. For further use, we denote $R^*_{\perp \alpha}=\varepsilon_{\alpha\lambda\beta\delta}\,e_{\perp}^{*\lambda} p^{\beta}\,n^{\delta}$. We use the following convention for the transverse euclidean polarization vectors $\eg^{\pm}$ as in eq.~(\ref{def_polL_polT})
\beq
\label{def_polT_2d}
 \eg^{\pm}=
\frac{1}{\sqrt{2}}\left(\mp1,- i\right)\,.
\eq
The DAs needed to parameterize the $\rho$ meson productions involved in the results of \cite{Anikin:2009bf} for the impact factors with $\Db=0$, 
 according 
 to \cite{Anikin:2009bf} are: the twist~2 DA $\varphi_1(y;\mu_F^2)$ associated to the production of the longitudinal $\rho$ meson, the two-parton twist~3 DAs $\varphi_1^T(y;\mu_F^2)$, $\varphi_A^T(y;\mu_F^2)$ and the three-parton twist~3 DAs $B(y_1,y_2;\mu_F^2)$ and $D(y_1,y_2;\mu_F^2)$ that parameterize the production of a transversely polarized $\rho$ meson. Following \cite{Anikin:2011sa,Besse:2012ia}, we also use the combinations,
\beqa
\label{defSM}
M(y_1,y_2;\mu_F^2)&=&\zeta^{V}_{3}(\mu_F^2) \, B(y_1,y_2;\mu_F^2)-\zeta^{A}_{3}(\mu_F^2) \, D(y_1,y_2;\mu_F^2)\,, \nonumber
\\
S(y_1,y_2;\mu_F^2)&=&\zeta^{V}_{3}(\mu_F^2) \, B(y_1,y_2;\mu_F^2)+\zeta^{A}_{3}(\mu_F^2) \,D(y_1,y_2;\mu_F^2)\,,
\eqa
where $\zeta^V_{3}(\mu_F^2)$ and $\zeta^{A}_{3}(\mu_F^2)$ are the dimensionless coupling constants
\beq
\label{def_zeta}
\zeta^{V}_{3}(\mu_F^2)=\frac{f^V_{3\rho}(\mu_F^2)}{f_{\rho}}\,, \qquad
\zeta^{A}_{3}(\mu_F^2)=\frac{f^A_{3\rho}(\mu_F^2)}{f_{\rho}}\,.
 \eq
 The scale $\mu_F$ is the factorization scale involved in the production of the $\rho$ meson, that we put equal to the renormalization scale of the evolution of the DAs. We recall in appendix~\ref{subsubsec:DA_LCCF} some basics features of the chiral even twist~2 and twist~3 DAs present in this approach; more details can be found on the DAs in ref.~\cite{Anikin:2009bf}. We recall also in appendix~\ref{subsec:DA_evolution} the dependence of the DAs on the collinear factorization scale $\mu_F$ which is driven by the renormalization evolution equations given in ref.~\cite{Ball:1998sk}.
 
We have recently shown in~\cite{Besse:2012ia} that these impact factors, expressed in the impact parameter space, read,
\bea
 \Phi^{\gamma^*_L \to \rho_L}&=&-\frac{1}{4}m_{\rho} f_{\rho} \int dy \int d^2 \rb \left(e^*\cdot n\right)\varphi_1(y;\mu_F^2) \tilde{H}_{q\qb}^{\sla p_1}(y,\rb,\kb)\,,
\label{IF_TF_L}
\eea
and
\bea
 \Phi^{\gamma^*_T \to \rho_T}&=&-\frac{1}{4}m_{\rho} f_{\rho} \int dy_2 \int d^2 \rb \left\{\varphi_3(y_2;\mu_F^2) \tilde{H}_{q\qb}^{\sla e^*_{\perp}}(y_2,\rb,\kb)+i \varphi^T_{1}(y_2;\mu_F^2)\left(\eb^*\cdot\rb\right) \tilde{H}_{q\qb}^{\sla p_1}(y_2,\rb,\kb)\right.\nn\\
&&\hspace{1.5cm}+\left.i\varphi_A(y_2;\mu_F^2) \tilde{H}_{q\qb}^{\,\sla \!R^*_{\perp} \gamma_5}(y_2,\rb,\kb)- \varphi^{T}_{A}(y_2;\mu_F^2)\left(\underline{R}^*\cdot\rb\right)\tilde{H}_{q\qb}^{\sla p_1 \gamma_5}(y_2,\rb,\kb)\right.\nn\\
&&\hspace{1.5cm}-\left.i\int_{0}^{y_2} dy_1 \int d^2\underline{r}' \,\left(\zeta^{V}_{3} B(y_1,y_2;\mu_F^2)\tilde{H}^{e^*_{\perp},\sla p_1}_{q \qb g}(y_1,y_2,\rb,\underline{r}',\kb)\right.\right.\nn\\
&&\hspace{4.5cm}\left.\left.+i \zeta^{A}_{3} D(y_1,y_2;\mu_F^2)\tilde{H}^{R^*_{\perp},\sla p_1 \gamma_5}_{q \qb g}(y_1,y_2,\rb,\underline{r}',\kb)\right)\right\}\,,
\label{IF_TF}
\eea
where we have respectively denoted in the case of two-parton exchange as $y$ and $\bar{y}=1-y$ the longitudinal momentum fractions of the quark and the antiquark, and in the three-parton exchange $y_1$, $\yb_2=1-y_2$ and $y_g=y_2-y_1$ the longitudinal momentum fractions of the quark, the antiquark and the gluon. The transverse displacement vectors of the color dipole configurations are denoted as $\rb$ for the interacting dipole (two- and three-parton Fock component) and $\rb'$ for the spectator dipole (present only in the three-parton Fock component). Note that $\left|\rb\right|$ in the case of the two-parton component is the transverse size of the quark anti-quark colorless pair. 
The functions denoted
\beq
\tilde{H}_{q\qb}^{\Gamma^{\mu}b_{\mu}} \equiv \tilde{H}_{q\qb}^{\Gamma^{\mu}} b_{\mu}\,, 
\quad
\tilde{H}^{c,\Gamma^{\mu}b_{\mu}}_{q\qb g}\equiv \tilde{H}_{q\qb g}^{\alpha,\Gamma^{\mu}}\,c_{\alpha}\,b_{\mu}\,,
\eq
are the Fourier transforms in the transverse plane of the two-parton component hard parts  $H$  and the three-parton component hard parts $H_\mu$ (illustrated in figure~\ref{Fig:NonFactorized}), projected on the appropriate Fierz structures $\Gamma^\mu$.

The computations of the hard parts lead to the following generic expressions,
\bea
\label{phiLpsi}
\Phi^{\gamma^{*}_L \rightarrow \rho_L}(\kb,Q,\mu_F^2)&=&\left(\frac{\delta^{ab}}{2}\right)\int dy \int d \rb\,\, \psi^{\gamma^*_L\to\rho_L}_{(q\qb)}(y,\rb;Q,\mu_F^2)\, \mathcal{A}(\rb,\kb)\,,\\
\Phi^{\gamma^{*}_T \rightarrow \rho_T}(\kb,Q,\mu_F^2)&=&\left(\frac{\delta^{ab}}{2}\right)\int dy \int d \rb\,\, \psi^{\gamma^*_T\to\rho_T}_{(q\qb)}(y,\rb;Q,\mu_F^2) \,\mathcal{A}(\rb,\kb)\\
&&\hspace{-1cm}+\left(\frac{\delta^{ab}}{2}\right)\int dy_2 \int dy_1 \int d \rb \,\, \psi^{\gamma^*_T\to\rho_T}_{(q\qb g)}(y_1,y_2,\rb;Q,\mu_F^2) \mathcal{A}(\rb,\kb)\,,\nn
\label{phiTpsi}
\eea
where the function 
\beq
\label{Dip_tGluon}
\mathcal{A}(\rb,\kb)=\frac{4\pi\alpha_s}{N_c} \left(1-\exp\left(i\kb\cdot\rb\right)\right)\left(1-\exp\left(-i\kb\cdot\rb\right)\right)\,,
\eq
is the scattering amplitude of a quark-antiquark color dipole with the two $t-$channel gluons, putting apart the color factor $Tr(t^a\,t^b)=\delta^{ab}/2$, with $a$ and $b$ color indices and $N_c$ the number of colors. The functions $\psi^{\gamma^*_L\to\rho_L}_{(q\qb)}$, $\psi^{\gamma^*_T\to\rho_T}_{(q\qb)}$, $\psi^{\gamma^*_T\to\rho_T}_{(q\qb g)}$ are respectively the amplitudes of production of a $\rho$ meson from a quark-antiquark (quark-antiquark gluon) system produced far upstream the target in the fluctuation of the virtual photon. These functions are computed up to twist 3 in the collinear aproximation in ref.~\cite{Besse:2012ia} and they contain information about the relevant color dipole system that interacts with the target. The functions $\psi^{\gamma^*_{L,T}\to\rho_{L,T}}_{(q\qb)}$ can be expressed in terms of the  virtual photon wave functions $\Psi_{(h,\bar h)}^{\gamma^*_{L,T}}$,
\bea
 \Psi_{(h,\bar h)}^{\gamma^*_{L}}(y,\rb;Q^2)&=&\delta_{\bar{h},-h}\frac{e}{2\pi}\sqrt{\frac{N_c}{\pi}} \frac{\mu^2}{Q} K_0( \mu \left|\rb\right| )\,,\\
 \Psi_{(h,\bar h)}^{\gamma^*_{T}\,(\lambda_{\gamma})}(y,\rb;Q^2)&=&\delta_{\bar{h},-h}\frac{i e}{2 \pi}\sqrt{\frac{N_c}{\pi}}(y \delta_{h,\lambda_{\gamma}}+\yb \delta_{h,-\lambda_{\gamma}})\frac{(\rb\cdot\eb^{(\lambda_{\gamma})})}{\left|\rb\right|}\mu  K_1(\mu \left|\rb\right| )\,,
\eea
where $h=\pm \frac{1}{2}$, $\bar{h}=\pm \frac{1}{2}$ denote respectively the helicities of the exchanged quark and anti-quark, and of the combinations of DAs of the $\rho$ meson $\phi^{\rho_{L,T}}_{ (h\bar{h})}$,
\bea
\phi^{\rho_L}_{(h,\bar{h})}(y;\mu_F^2)&=&\delta_{\bar{h},-h}\sqrt{\frac{\pi}{4N_c}}(e_{L}^*\cdot n)\varphi_1(y;\mu_F^2)\,,\\
\phi_{(h,\bar{h})}^{\rho_T,\,(\lambda_{\rho})}(y,\rb;\mu_F^2)&=&-\delta_{\bar{h},-h}i\sqrt{\frac{\pi}{4 N_c}} (\eb^{(\lambda_{\rho})*}\cdot \rb)\nn\\
&\times & \left(\varphi_A^{T}(y;\mu_F^2)+(\delta_{h,\lambda_{\rho}}-\delta_{h,-\lambda_{\rho}})\varphi_1^{T}(y;\mu_F^2)\right)\,.
\eea
In the two-parton approximation, these functions $\phi^{\rho_L}_{(h,\bar{h})}(y;\mu_F^2)$ and $\phi_{(h,\bar{h})}^{\rho_T,\,(\lambda_{\rho})}(y,\rb;\mu_F^2)$ parameterize the moments of the wave functions of the $\rho$ meson, i.e. the first terms of the Taylor expansion of the wave functions at small $\rb$. Note that in the approach of ref.~\cite{Ivanov:1998gk} $\phi^{\rho_L}_{(h,\bar{h})}(y;\mu_F^2)$ and $\phi_{(h,\bar{h})}^{\rho_T,\,(\lambda_{\rho})}(y,\rb;\mu_F^2)$ are replaced by the Taylor expansion for small $\rb$ of the modeled wave functions of the vector mesons.
Finally, the functions $\psi^{\gamma^*_L\to\rho_L}_{(q\qb)}$ and $\psi^{\gamma^*_T\to\rho_T}_{(q\qb)}$ with two-parton components read
\bea
\label{eqlL}
\psi^{\gamma^*_L\to\rho_L}_{(q\qb)}(y,\rb;Q,\mu_F^2)&=&\frac{m_{\rho}f_{\rho}}{\sqrt{2}}\sum_{(h,\bar{h})}  \phi^{\rho_{L}}_{ (h\bar{h})}(y;\mu_F^2)
\, \Psi_{(h,\bar{h})}^{\gamma^*_{L}}(y,\rb;Q^2) \,,\\
\label{eqlT}
\psi^{\gamma^*_T\to\rho_T}_{(q\qb)}(y,\rb;Q,\mu_F^2)&=&\frac{m_{\rho}f_{\rho}}{\sqrt{2}}\sum_{(h,\bar{h})} \phi^{\rho_T,\,(\lambda_{\rho})}_{(h,\bar{h})}(y;\mu_F^2)
\, \Psi_{ (h,\bar{h})}^{\gamma^*_{T}\,(\lambda{\gamma})}(y,\rb;Q^2) \,.
\eea
The function $\psi^{\gamma^*_T\to\rho_T}_{(q\qb g)}$ with three-parton components reads
\bea
\label{eqlT3}
\psi^{\gamma^*_T\to\rho_T}_{(q\qb g)}(y,\rb;Q,\mu_F^2)&=&\frac{m_{\rho}f_{\rho}}{\sqrt{2}}\left[\left(\sqrt{\frac{\pi}{4 N_c}}\frac{S(y_1,y_2;\mu_F^2)}{2}\right) \mathcal{F}^{\gamma^*_T }(y_1,y_2,\rb;Q)\right.\nn\\
&&\hspace{1.5cm}-\left.\left(\sqrt{\frac{\pi}{4 N_c}}\frac{M(y_1,y_2;\mu_F^2)}{2}\right) \mathcal{F}^{\gamma^*_T}(\yb_2,\yb_1,\rb;Q)\right]  \,,
\eea
where the function $\mathcal{F}^{\gamma^*_T}$ describes the fluctuation of the transversely polarized photon into a quark-antiquark-gluon color singlet. The function $\mathcal{F}^{\gamma^*_T}$ can be expressed in terms of the longitudinally polarized photon
wave function 
\beq
\Psi^{\gamma^*_{L}}(\mu_i,\rb;Q)= \sum_{(h,\bar{h})} \Psi_{(h,\bar{h})}^{\gamma^*_{L}}\equiv 2\frac{e}{2\pi}\sqrt{\frac{N_c}{\pi}}\frac{\mu_i^2}{Q}K_0(\mu_i\left|\rb\right|)\,,
\eq
as
\bea
\label{F3part}
&& \mathcal{F}^{\gamma^*_T}(y_1,y_2,\rb;Q)=\frac{1}{2} \left\{2\,\left[\frac{\Psi^{\gamma^*_{L}}(\mu_1,\rb;Q)}{\yb_1 Q}\right]\right.\nonumber\\
&&\left. +\frac{N_C}{C_F}\left[\,\frac{\Psi^{\gamma^*_{L}}(\mu_{\qb g},\rb;Q)}{\yb_1 Q}+\left(\frac{y_2 \,\yb_1}{\yb_2 \,y_1}\right)\times \left(\frac{\Psi^{\gamma^*_{L}}(\mu_2,\rb;Q)}{\yb_1 Q}-\frac{\Psi^{\gamma^*_{L}}(\mu_{\qb g},\rb;Q)}{\yb_1 Q}
\right)\right]\right.\nonumber\\
&&+\left.\left(\frac{N_C}{C_F}-2\right)\left[\left(\frac{\Psi^{\gamma^*_{L}}(\mu_1,\rb;Q)}{y_g Q}-\frac{\Psi^{\gamma^*_{L}}(\mu_{q\qb},\rb;Q)}{y_g Q}
\right)\right.\right.\nn\\
&&\left.\left.\hspace{2.5cm}+\frac{y_2}{\yb_2}\left(\frac{\Psi^{\gamma^*_{L}}(\mu_2,\rb;Q)}{y_g Q}-\frac{\Psi^{\gamma^*_{L}}(\mu_{q\qb},\rb;Q)}{y_g Q}\right)\right]\right\}\,,
\eea
with 
\bea
\label{defmuGEN}
\mu_1^2=y_1\yb_1Q^2\,,&\quad& \mu_2^2=y_2\yb_2Q^2\,,\\
\mu_{q g}^2=\frac{y_1y_g}{y_1+y_g}Q^2\,,\quad \mu_{\qb g}^2&=&\frac{\yb_2y_g}{\yb_2+y_g}Q^2\,,\quad \mu_{q \qb}^2=\frac{y_1\yb_2}{y_1+\yb_2}Q^2\,,\nn
\eea
and $C_F=\frac{N_c^2-1}{2N_c}$. 
Note, that in the large $N_C$ limit $\mathcal{F}^{\gamma^*_T}$ simplifies,
\bea
\label{F3part-Nc-large}
&& \mathcal{F}^{\gamma^*_T}(y_1,y_2,\rb;Q) \xrightarrow[N_C\to\infty]{} \frac{1}{\yb_1 y_1 \yb_2 Q} \\
&&\times \left\{y_1\yb_2 \Psi^{\gamma^*_{L}}(\mu_1,\rb;Q)+y_2 \yb_1\Psi^{\gamma^*_{L}}(\mu_2,\rb;Q) -y_g\Psi^{\gamma^*_{L}}(\mu_{\qb g},\rb;Q)\right\}\,.\nn
\eea
\subsection{From impact factors to helicity amplitudes and polarized cross-sections}
\label{subsec:HAandPC}
According to eq.~(\ref{conv-phi-F1}), the helicity amplitudes read,
\beq
\frac{T_{\lambda_{\rho},\lambda_{\gamma}}}{s}
=\frac{\delta^{ab}}2 \int \frac{d^2 \kb}{\kb^4} \Phi_{ab}^{\gamma^{*}_{\lambda_{\gamma}} \rightarrow \rho_{\lambda_{\rho}}}(\kb,Q,\mu_F^2) \mathcal{F}(x,\kb)\,.
\label{conv-phi-F}
\eq
Inserting the expressions for the impact factor $\Phi^{\gamma^{*}_{\lambda_{\gamma}} \rightarrow \rho_{\lambda_{\rho}}}$ of eqs.~(\ref{phiLpsi}, \ref{phiTpsi}), one gets 
\bea
\label{T00Lpsi}
\frac{T_{00}}{s}&=&\int dy \int d \rb\, \psi^{\gamma^*_L\to\rho_L}_{(q\qb)}(y,\rb;Q,\mu_F^2) \,\hat{\sigma}(x,\rb)\,,\\
\label{T11Tpsi}\frac{T_{11}}{s}&=&\int dy \int d \rb\, \psi^{\gamma^*_T\to\rho_T}_{(q\qb)}(y,\rb;Q,\mu_F^2)\, \hat{\sigma}(x,\rb)\\
&&\hspace{-1cm}+\int dy_2 \int dy_1 \int d \rb \, \psi^{\gamma^*_T\to\rho_T}_{(q\qb g)}(y_1,y_2,\rb;Q,\mu_F^2)\, \hat{\sigma}(x,\rb)\,,\nn
\eea
where we have identified the dipole cross-section as defined in ref.~\cite{Forshaw:1997dc}
\bea
\hat{\sigma}(x,\rb)&=&\frac{N_c^2-1}{4} \int \frac{d^2\kb}{\kb^4}\mathcal{F}(x,\kb)\mathcal{A}(\kb,\rb) \,.
\eea
Note that we can separate the $T_{11}$ as the WW contribution and the genuine contribution, 
\bea
\label{T11WW}
\frac{T^{WW}_{11}}{s}&=&\int dy \int d \rb\, \psi^{\gamma^*_T\to\rho_T,\,WW}_{(q\qb)}(y,\rb;Q,\mu_F^2)\, \hat{\sigma}(x,\rb)\,,\\
\frac{T^{gen}_{11}}{s}&=&\int dy \int d \rb\, \psi^{\gamma^*_T\to\rho_T,\,gen}_{(q\qb)}(y,\rb;Q,\mu_F^2)\, \hat{\sigma}(x,\rb)\\
&&\hspace{-1cm}+\int dy_2 \int dy_1 \int d \rb \, \psi^{\gamma^*_T\to\rho_T}_{(q\qb g)}(y_1,y_2,\rb;Q,\mu_F^2)\, \hat{\sigma}(x,\rb)\,,\nn
\label{T11Genuine}
\eea
The formulas~(\ref{T00Lpsi}, \ref{T11Tpsi}) allow us to combine various models of the scattering amplitude of a dipole on a nucleon with the results obtained by twist expansion of the $\gamma^*\to \rho$ impact factor. At $t=t_{\rm min}$ the  contribution to the cross-sections $\sigma_{L}$ and $\sigma_{T}$ are respectively coming from the helicity amplitudes  $T_{00}$ and $T_{11}$,
\bea
\label{Dsigma00}
\frac{d \sigma_L}{d t}(t=0) &=& \frac{\left|T_{00}(s,t=0)\right|^2 }{16\pi s^2}\,,\\
\frac{d \sigma_T}{d t}(t=0) &=& \frac{\left|T_{11}(s,t=0)\right|^2 }{16\pi s^2}\,.
\label{Dsigma11}
\eea
The $t-$dependency is expected to be governed by non-perturbative effects of the nucleon, which can be phenomenologically parameterized by
 an exponential dependence of the differential cross-section 
\beq
\label{t-dependence}
\frac{d \sigma_{L,T}}{d t}(t)=e^{-b(Q^2) t}\,\frac{d \sigma_{L,T}}{d t}(t=0)\,.
\eq
This results in the polarized cross-sections
\bea
\label{sigmaLprediction}
\sigma_L &=&\frac{1}{b(Q^2)}\frac{\left|T_{00} (s,t=0)\right|^2 }{16\pi s^2}\,,\\
\label{sigmaTprediction}
\sigma_T &=&\frac{1}{b(Q^2)}\frac{\left|T_{11} (s,t=0)\right|^2 }{16\pi s^2}\,.
\eea
The $b(Q^2)$ slope  has been measured by ZEUS and H1. We will use here quadratic fits of the $b(Q^2)$ slope data of ref.~\cite{Aaron:2009xp} to determine the cross-section.

In the following section, we will briefly present the dipole models 
we shall use to compare our predictions with HERA data.
Note that we could in principle use a dipole model taking into account  skewness effects. 
This skewness dependence can be implemented along approaches of refs.~\cite{Shuvaev:1999ce,Martin:1999wb},
 but this 
 subleading physical effects will be neglected in the present study.

\section{Dipole models}
\label{sec:Dipole models}

In the dipole picture, the DIS cross-section reads \cite{Mueller:1989st,Nikolaev:1990ja}
\bea
\label{DIS-dipolepicture}
\sigma^{\gamma^*p}_{L,T}&=&\int d^2\rb  \int dy \sum^{N_f}_{\text{f}}\left|\Psi^{\gamma^*_{L,T}}_f(y,\rb;Q)\right|^2\,\hat{\sigma}(x,\rb)\,,
\eea
with
\bea
\label{Sigma-b-dep}
\hat{\sigma}(x,\rb) &=&\int d^2 \underline{b}\, \frac{d \hat{\sigma}_{q\qb}}{d^2\underline{b}}=2 \int d^2 \underline{b}\,\, \mathcal{N}(x,\rb,\underline{b})\,.
\eea
The low-$x$ saturation dynamics of the nucleon target was first introduced in refs.~\cite{GolecBiernat:1998js,GolecBiernat:1999qd} by Golec-Biernat and W\"usthoff (GBW) model to describe the inclusive and diffractive structure functions of DIS, which inspired many phenomenological descriptions of DIS HERA data \cite{Gotsman:2002yy,Bartels:2002cj,Iancu:2003ge,Kowalski:2003hm,Albacete:2005ef,Kowalski:2006hc,Goncalves:2006yt}. In this model the dipole cross-section reads
\beq
\label{sigmaGBW}
\hat{\sigma}(x,r)=\sigma_0\left\{1-\exp\left(-\frac{r^2}{4R_0^2(x)}\right)\right\}\,,
\eq
where 
\beq
\label{R0}
R_0^2(x)=\frac{1}{\text{GeV}^2} \left(\frac{x}{x_0}\right)^{\lambda_p}\,,
\eq
and it involves three independent parameters $\left\{\sigma_0,x_0,\lambda_p\right\}$. One can see from eq.~(\ref{DIS-dipolepicture}) that since the wave functions are peaked at $r\sim\frac{1}Q$, the domain in which the saturation effects are significant is given by 
\beq
\label{Qsat}
Q^2\lesssim\frac{1}{R_0^2(x)}\equiv Q_S^2(x)\,.
\eq
To make contact with photoproduction, it is customary \cite{GolecBiernat:1998js} 
to make the following modification of the definition of
 the Bjorken variable $x$  
\beq
\label{redefxBj}
x \to  x\left(1+\frac{4\, m_f^2}{Q^2}\right)=\frac{Q^2}{W^2+Q^2}\left(1+\frac{4\, m_f^2}{Q^2}\right) 
\xrightarrow[Q^2 \to 0]{} \frac{4 \, m_f^2}{W^2}\,,
\eq
where $m_f$ is an effective quark mass which depends on the flavour $f$ and of the model used to fit the data. 
This modification is adopted in the following.

 In the GBW saturation model, light quark masses are taken to be $0.14\;$GeV in order to get a good fit of the photoproduction region. The inclusion of the charm contribution, with $m_c=1.5\;$GeV has also been performed in ref.~\cite{GolecBiernat:1998js}. 
 The normalisation $\sigma_0$ results from the integration over the impact parameter $b$, assuming that the $b-$dependence of the dipole amplitude factorises as $\mathcal{N}(x,\rb,\underline{b})=T(\underline{b})\, \mathcal{N}(x,r)$, where $T(b)$ is related to the density of gluon within the target, leading to
\bea
\label{Sigma-b-dep2}
\hat{\sigma}(x,\rb) &=&2 \int d^2\underline{b} \,T(\underline{b})\, \mathcal{N}(x,r)=\sigma_0 \, \mathcal{N}(x,r)\,.
\eea
 This model provides a good description of inclusive and diffractive structure functions for the values of the parameters presented in table~\ref{table:GBWfits}.
 \begin{table}[h!]
 	\centering
  \begin{tabular}{|l|c|c|c|c|c|c|r|}
 \hline
  Fits   &  $\sigma_0$ (mb) & $\lambda_p$  & $x_0$ 
   \\ \hline 
 No charm  &  23.03 &  0.288  & $3.04 \times 10^{-4}$  
 \\ \hline
With charm  & 29.12  & 0.277   & $0.41\times 10^{-4}$ 
\\ \hline
 \end{tabular}
 \caption{Values of the parameters entering the GBW dipole cross-section.}
 \label{table:GBWfits}
 \end{table}

The small $x$ evolution of the dipole cross-section can be modeled as in the refs.~\cite{GolecBiernat:1998js},
or computed numerically from the running coupling  Balitsky-Kovchegov (rcBK) equation. Indeed, the $x-$dependence is driven at small $x$ by perturbative non-linear equations, the Jalilian-Marian-Iancu-McLerran-Weigert-Leonidov-Kovner (JIMWLK) equation \cite{JalilianMarian:1997gr,JalilianMarian:1997dw,Kovner:2000pt,Weigert:2000gi,Iancu:2000hn,Ferreiro:2001qy} and the Balitsky-Kovchegov (BK) equation \cite{Balitsky:1995ub,Kovchegov:1999yj}. The solutions of the two equations are not significantly different and the BK equation is simpler to solve numerically than the JIMWLK equation. The LO-BK equation cannot be used to obtain the low $x$ evolution as it predicts a growth of the saturation scale much faster than the growth expected from the anlysis based on phenomenological models. It was shown in \cite{Albacete:2007yr,Albacete:2007sm} that taking into account only the running coupling corrections of the evolution kernel allows  to get the main higher order contributions, and to solve the discrepancy between the growths of the saturation scale. 
 A numerical solution of the running coupling BK (rcBK) equation was obtained in refs.~\cite{Albacete:2009fh,Albacete:2010sy} based on initial conditions inspired by the GBW model \cite{GolecBiernat:1998js} and the McLerran-Venugopalan (MV) model \cite{McLerran:1997fk}. We will denote these numerical solutions for the dipole scattering amplitudes as the Albacete-Armesto-Milhano-Quiroga-Salgado (AAMQS)-model.
\bea
\label{GBWinitial}
\mathcal{N}^{GBW}(x_0,r)&=&\sigma_0^{GBW}\left\{ 1-\exp \left[-\left(\frac{r^2 Q^2_{s\,0}}{4 }\right)^{\gamma}\right]\right\}\,,\\
\mathcal{N}^{MV}(x_0,r)&=&\sigma_0^{MV}\left\{1-\exp\left[-\left(\frac{r^2Q^2_{s\,0}}{4 }\right)^{\gamma}\,\ln\left(\frac{r}{\Lambda_{QCD}}+e\right)\right]\right\}\,,
\label{MVinitial}
\eea
with $x_0=0.01$ and $Q_{s\,0}$ the initial saturation scale at $x=x_0$, and $\gamma$ the anomalous dimension. The coupling constant in the evolution kernel of the rcBK equation depends on the number of active quark flavors $n_f$,
\bea
\alpha_{s,n_f}(r^2)&=&\frac{4\pi}{\beta_{0,n_f}\,\ln\left[\frac{4C^2}{r^2 \Lambda^2_{n_f}}\right]}\,,
\eea
where $\beta_{0,n_f}=11-\frac{2}{3}n_f$, $\Lambda_{n_f}$ is the QCD scale and $C$ is one of the free parameters of the model. As usual, the scales $\Lambda_{n_f}$ are determined by the matching condition $\alpha_{s,n_f-1}(r_{\star}^2)=\alpha_{s,n_f}(r_{\star}^2)$ at $r_{\star}^2=4C^2/m_f^2$ and an experimental value of $\alpha_s$. In eq.~(\ref{MVinitial}), the scale $\Lambda_{QCD}$ is identified with $\Lambda_3$.

The parameters are fitted to the experimental data for the inclusive structure function of DIS
\beq
F_2(x,Q^2)=\frac{Q^2}{4\pi^2 \alpha_{em}}\left(\sigma_T+\sigma_L\right)\,,
\eq
and give a good description for the data of the longitudinal structure function
\beq
F_L(x,Q^2)=\frac{Q^2}{4\pi^2 \alpha_{em}}\sigma_L\,.
\eq
Two sets of dipole cross-sections are available for each of the initial conditions. The first set of dipole cross-sections respectively denoted as (a) and (e) for the GBW and the MV initial conditions, are fitted to data by taking only into account the light quarks $u,d,s\,.$
The second set, respectively denoted as (b) and (f) for the GBW and the MV initial conditions, are fitted taking into account the dipole cross-section of the light quarks $\sigma_0 \, \mathcal{N}^{\text{l.}}(x,\rb)$ and of the heavy quarks $\sigma_0^{\rm h} \mathcal{N}^{\text{h.}}(x,\rb)\,.$
The normalization of the dipole cross-section $\sigma_0^{{\rm h.}}$ for the charm and the bottom quarks are assumed to be equal. Thus the cross-sections read
\bea
\label{SigmaSetae}
\sigma^{\gamma^*p}_{L,T;\,\text{set(a),(e)}}&=&\sigma_0\sum_{\text{f=u,d,s}} \int d^2\rb  \int dy \left|\Psi^{\gamma^*_{L,T}}_f(y,\rb;Q,m_f,e_f)\right|^2\,\mathcal{N}^{\text{l.}}(x,\rb)\,,\\
\label{SigmaSetbf}
\sigma^{\gamma^*p}_{L,T,\,\text{set(b),(f)}}&=&\sigma_0^{l.}\sum_{\text{f=u,d,s}} \int d^2\rb  \int dy \left|\Psi^{\gamma^*_{L,T}}_f(y,\rb;Q,m_f,e_f)\right|^2\,\mathcal{N}^{\text{l.}}(x,\rb)\nn\\
&+&\sigma_0^{h.}\sum_{\text{f=c,b}} \int d^2\rb  \int dy \left|\Psi^{\gamma^*_{L,T}}_f(y,\rb;Q,m_f,e_f)\right|^2\,\mathcal{N}^{\text{h.}}(x,\rb)\,.
\eea 
In the following parts we will focus mostly on the sets (a) and (b) using the GBW initial condition respectively denoted as AAMQSa and AAMQSb, as the effects due to different initial conditions do not involve significant changes in the numerical results of present study. 
 We present in Tabs.~\ref{tableAAMQSfits} and \ref{tableAAMQSfits2} values of the parameters of the fits obtained in ref.~\cite{Albacete:2010sy}.
\begin{table}[h!]
 	\centering
  \begin{tabular}{|l|c|c|c|c|r|}
 \hline
  Fits  & $Q_{s0}^2$ & $\sigma_0$ (mb) & $\gamma$ & $C$ & $\chi^2/N_{df}$ \\ \hline 
 (a) & 0.241 & 32.357 & 0.971 & 2.46 & 1.226 \\ \hline
(e) & 0.165 & 32.895 & 1.135 & 2.52 & 1.171 \\ \hline
 \end{tabular}
 \caption{Values of the parameters entering the AAMQS sets (a) and (e) dipole cross-sections.}
 \label{tableAAMQSfits}
 \end{table}
 \begin{table}[h!]
 	\centering
  \begin{tabular}{|l|c|c|c|c|c|c|c|r|}
 \hline
  Fits  & $Q_{s0}^2$ & $Q_{s0}^{(c,b)\,2}$ & $\sigma_0^{{\rm l.}}$ (mb) & $\sigma_0^{{\rm h.}}$ (mb) & $\gamma$ & $\gamma^{(c,b)}$ &  $C$ & $\chi^2/N_{df}$ \\ \hline 
 (b) & 0.2386& 0.2329 & 35.465 & 18.430 & 1.263 & 0.883 & 3.902 & 1.231 \\ \hline
  (f) & 0.1687& 0.1417 & 35.449 & 19.066 & 1.369 & 1.035 & 4.079 & 1.244 \\ \hline
 \end{tabular}
 \caption{Values of the parameters entering the AAMQS sets (b) and (f) dipole cross-sections.}
 \label{tableAAMQSfits2}
 \end{table}

Note that the models of dipole cross-section we use here involve only a quark anti-quark intermediate state between the initial and final wave functions used to describe DIS process while the impact factor $\gamma^*_T\to\rho_T$ involves also $q\qb g$ partonic state. The error induced by this approximation should be subleading compare to higher twist corrections or the choice of $\mu_F$, but still it is difficult to appreciate quantitatively how much it could impact on the predictions.

For completeness, we will also display predictions using the Gunion and Soper model (GS-model) \cite{Gunion:1976iy}. This model was used in our first phenomenological study of ref.~\cite{Anikin:2011sa}, and it 
assumes that the hadron impact factor takes the form
\begin{equation}
\label{ProtonIF}
\Phi^{N \to N}(\underline{k},\Db;M^2)= A \, \delta_{ab}\left[\frac{1}{M^2+(\frac{\Db}{2})^2}-\frac{1}{M^2+(\underline{k}-\frac{\Db}{2})^2}\right]\,,
\end{equation}
where $A$ and $M$ are free parameters that correspond to the soft scales of the proton-proton impact factor.
As it was discussed in~\cite{Anikin:2011sa}, the ratios of helicity amplitudes are well described for $M \simeq 1\;$GeV and the result is not very sensitive to this parameter. Note that this impact factor indeed vanishes when $\underline{k} \to 0$ or $\Db -\underline{k} \to 0$ in a minimal way.
 Such a model was the basis of the dipole approach of high energy scattering  \cite{Mueller:1993rr} and used successfully for describing deep inelastic scattering (DIS) at small $x$ \cite{Navelet:1996jx}. 
The dipole-proton scattering amplitude in the two-gluon exchange approximation, computed in details in appendix \ref{subsec:DipProtonGS}, resulting from the GS-model for the proton impact factor is:
\bea
\mathcal{N}(\rb,M)&=&\int \frac{d^2 \kb}{(2\pi)^2} \frac{1}{(\kb^2)^2}\, A\delta_{ab}\left(\frac{1}{M^2} - \frac{1}{M^2+\kb^2}\right) \left(1 - e^{i \kb \cdot \rb}\right) \left(1 - e^{-i \kb \cdot \rb}\right)\\
&=& \frac{A \delta_{ab}}{\pi M^4} \left(\gamma + \ln \frac{M \, r}2 + K_0(M  \, r)\right)\,.
\label{DipProtonGS}
\eea 


\section{Comparison with the HERA data} 
\label{sec:HERADATA}
Our main results are the polarized cross-sections $\sigma_T$ and $\sigma_L$ of the process (\ref{process}), that we compare with the data of the H1 collaboration \cite{Aaron:2009xp}.

\begin{figure}[h!]
\hspace*{\fill}
\begin{tabular}[p]{c}
\subfigure[$\sigma_T$ vs H1 data]{\epsfxsize=12 cm\epsfbox{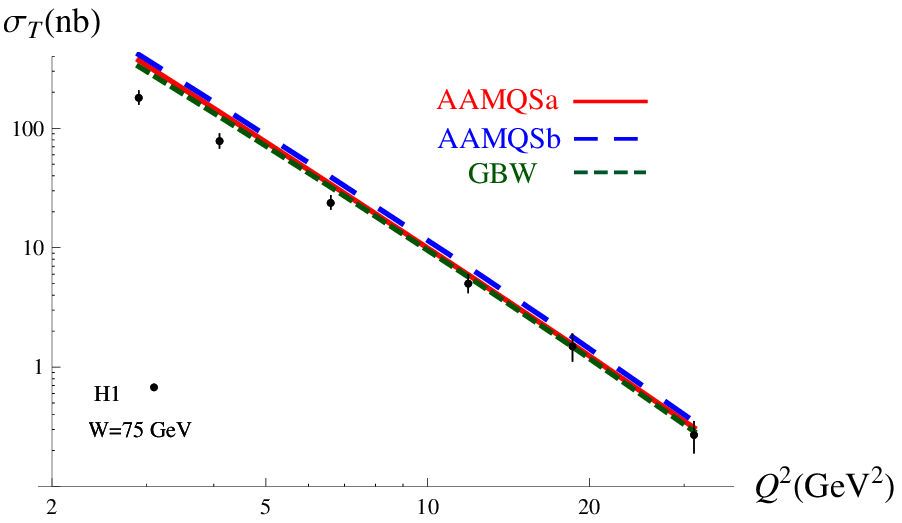}\label{Fig41}}\\
 \subfigure[$\sigma_L$ vs H1 data]{\epsfxsize=12 cm \epsfbox{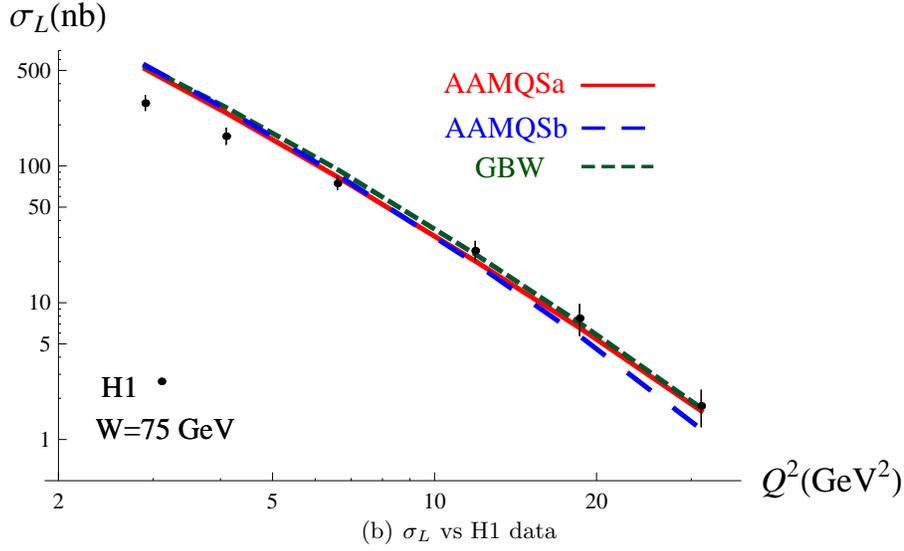}\label{Fig42}}
\end{tabular}
\hspace*{\fill}
\caption{Predictions for $\sigma_T$  and $\sigma_L$ vs $Q^2$, for $W=75\;$GeV, using the AAMQSa (red solid line), AAMQSb (blue large dashed line) and GBW (green dashed line) models compared to the data of H1\cite{Aaron:2009xp}.}
\label{FigAAGBW}
\end{figure}

 In the following plots, experimental errors are taken to be the quadratic sum of statistical and systematical errors. 
The collinear factorization scale $\mu_F$ only appears in the scattering amplitudes through the DAs and the coupling constants and unless specified we will assume that $\mu_F$ depends on the virtuality $Q^2$ as
\beq
\label{muF}
\mu_F^2(Q^2)=\frac{Q^2+M_{\rho}^2}{4} \xrightarrow[]{Q^2>>M_{\rho}^2} \frac{Q^2}4\,.
\eq

The figures~\ref{Fig41}~and~\ref{Fig42} show the full twist~3 predictions respectively for $\sigma_T$ and $\sigma_L$, using the different dipole models. 
The results of the predictions for $\sigma_T$ (figure~\ref{Fig41}) and $\sigma_L$ (figure~\ref{Fig42}) are in agreement with the data for values of $Q^2$ larger than approximately 7\;GeV$^2$ for $\sigma_T$ 
 and $7$\;GeV$^2$ for $\sigma_L$ depending on the considered dipole model. 
 The results based on the AAMQSa model are giving a slightly better description than with the AAMQSb although both sets give good quality fits of DIS data. 
  The results obtained by GBW model are close to the one obtained by the AAMQSa model; both are in good agreement with the data for $Q^2$ above 7\;GeV$^2$. Let us insist on the fact that the agreement for $Q^2\geq 7\;$GeV$^2$ of our predictions with data is non-trivial since all free parameters of the dipole cross-section (normalization, $R_{0}(x)$, etc...) are completely fixed by the fit of DIS data, while on the $\rho$ side the normalizations are given by decay constants obtained from QCD sum rules.

Keeping in mind that the results are not too sensitive to the precise choice of the dipole model, below we will focus on the predictions of the AAMQSa model. Some results of the GBW, AAMQSb and GS models are presented in appendix~\ref{subsec:GBW-AAb}.

\begin{figure}[h!]
\hspace*{\fill}
\begin{tabular}[p]{c}
\subfigure[AS (purple dashed line), WW (blue long dashed line) and Total (red solid line) contributions to $\sigma_T$.]{\epsfxsize=12 cm\epsfbox{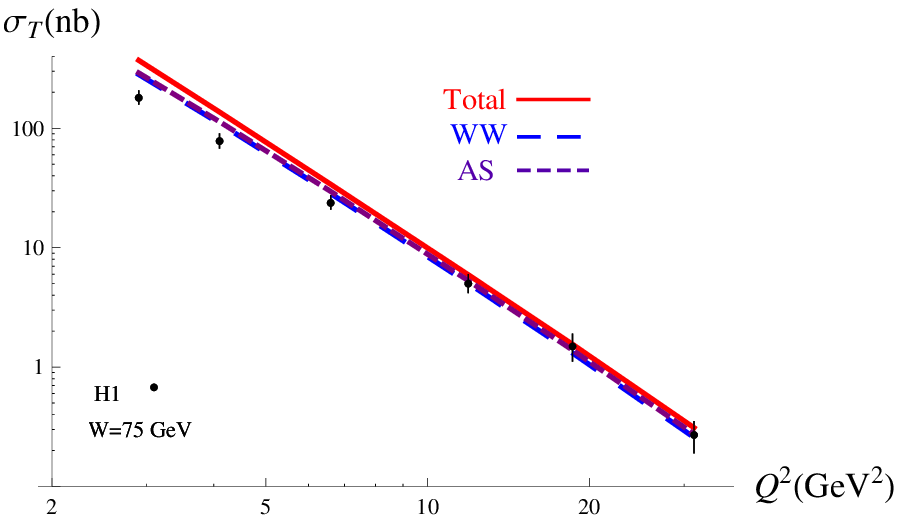}\label{sigT}}\\
 \subfigure[AS (purple dashed line) and Total (red solid line) contributions to $\sigma_L$.]{\epsfxsize=12 cm\epsfbox{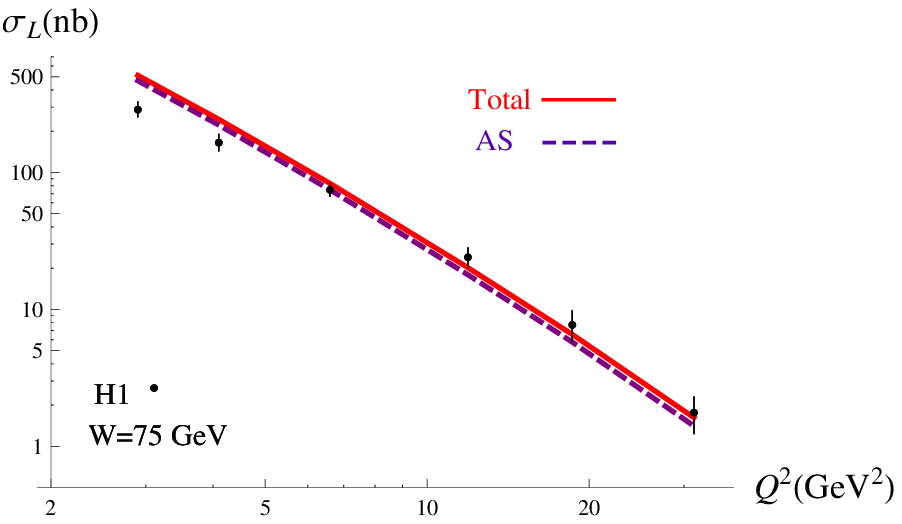}\label{sigL}}
\end{tabular}
\hspace*{\fill}
\caption{Predictions for $\sigma_T$  and $\sigma_L$ vs $Q^2$, for $W=75\;$GeV, using the AAMQSa-model, compared to the data of H1\cite{Aaron:2009xp}.}
\label{FigAAa1}
\end{figure} 
In figure~\ref{FigAAa1}, we show separately three different contributions:
\bei
\item the full twist~3 (Total) contribution, involving both the WW and the genuine solutions of the DAs. 
\item the WW contribution, only involving the WW solutions of the DAs. 
\item the asymptotic (AS) contribution (for  $\mu_F\to\infty$), involving the asymptotic solutions of the DAs. In this limit the genuine contribution vanishes and the WW DAs can be expressed as functions of the asymptotic DA $\varphi_1(y)=6y\yb$.
\ei

The results of the predictions for $\sigma_T$ (figure~\ref{sigT}) and $\sigma_L$ (figure~\ref{sigL}) are in agreement with the data for values of $Q^2$ respectively larger than $Q^{2 \,{\rm min}}_T\sim $ 6.5\;GeV$^2$ and $Q^{2\,{\rm min}}_L\sim 5$\;GeV$^2$ which confirms that the amplitude factorizes into a universal colour dipole scattering amplitude and that the truncated twist expansion of the $\rho$ meson soft part is justified.

The two scales $Q^{2\,\text{min}}_{T}$ and $Q^{2\,\text{min}}_L$ are close to each other. We interpret this fact as an indication that the discrepancy between data and our predictions at low $Q^2$ are mainly due to the higher twist contributions to the impact factor from the meson structure rather than an effect of the saturation dynamics of the nucleon which should be well described at this scale by the saturation models. Let us mention, again that these saturation models are known to fit very well inclusive DIS as well as diffractive DIS data (for GBW) at these low $Q^2$ values. 

The saturation scale, given by $Q_{S}=1/R_{0}(x)$ is of order $1\;$GeV in the kinematics of HERA. 
 Since our predictions are only consistent with data in the region $Q^2>Q^{2\,\text{min}}_{L,T}>Q^2_{S}$, 
this limitation do not allow us to access the 
 domain $Q^2_{S}\gtrsim Q^2$ where saturation effects can be essential.

The predictions are dominated by the WW-contribution and are not very sensitive to the choice of the collinear factorization scale. 
This will be further discussed in section~\ref{sec:distribution}.

\begin{figure}[h!]
	\centering
		\includegraphics{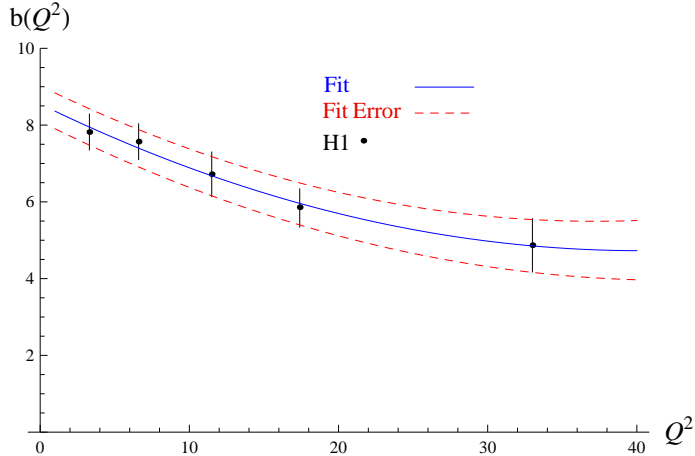}
\caption{Quadratic fits of the $b-$slope H1 data.}
\label{Figbslope}
\end{figure}
\begin{figure}[h!]
\hspace*{\fill}
\begin{tabular}[p]{c}
\subfigure[AS (purple) and Total (red) contributions to $\sigma_T$.]{\epsfxsize=12 cm\epsfbox{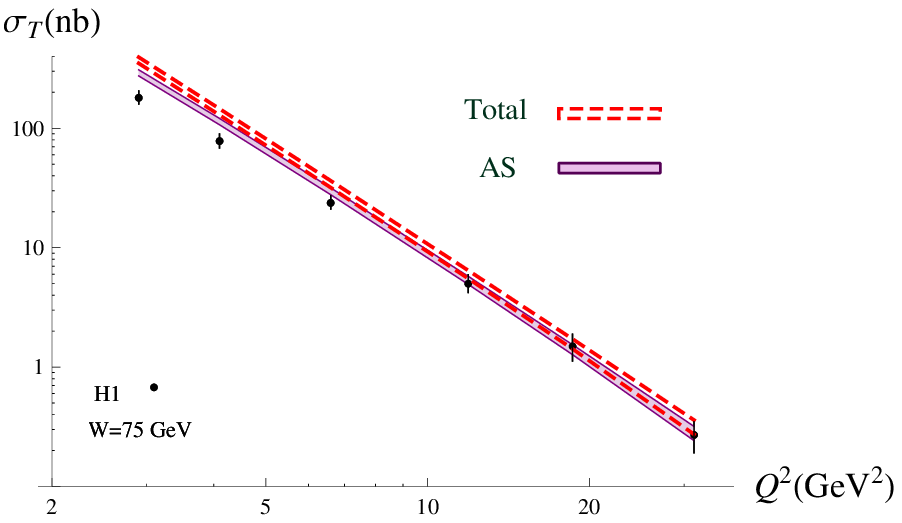}\label{deltasigT}}\\
 \subfigure[AS (purple) and Total (red) contributions to $\sigma_L$.]{\epsfxsize=12 cm\epsfbox{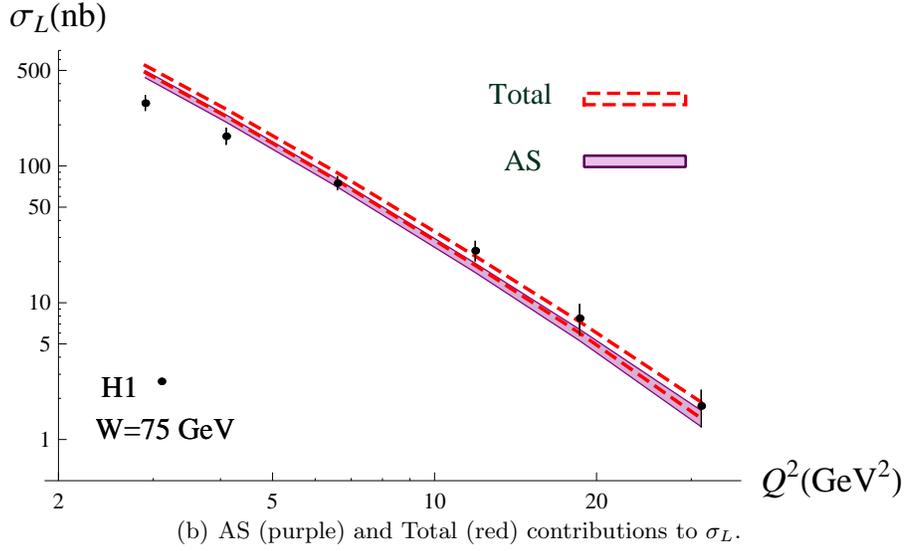}\label{deltasigL}}
\end{tabular}
\hspace*{\fill}
\caption{Full twist~3 and asymptotic predictions with the $b-$slope uncertainty, using AAMQSa model.}
\label{Fig112}
\end{figure}
An estimation of the error on the cross-sections caused by the error bars on the $b-$slope measurements is obtained by fitting the upper and lower bounds of the $b-$slope as shown in figure~\ref{Figbslope}, and then by computing the predictions based on these fits as shown in figure~\ref{Fig112}. Note that we have assumed that the longitudinal $b_L$ and the transverse $b_T$ slopes are equal to the $b-$slope of the total cross-section. This assumption is supported by H1 data where the measurements of the difference $b_L-b_T$ for $Q^2=3.3\;$GeV$^2$ and $Q^2=8.6\;$GeV$^2$ are much smaller than the $b-$slope value. Let us also emphasize that in this approach we compute the polarized differential cross-sections in the limit $t=t_{\rm min}\approx 0$ where only the $s$-channel helicity conserving (SCHC) amplitudes $T_{00}$ and $T_{11}$ are non-zero. The contributions of other helicity amplitudes are encoded in the phenomenological $t-$dependence given in eq.~(\ref{t-dependence}) and
it turns out that data for the total differential cross-section are dominated by a $t$-region of very small values,  with a typical spread given by the scale $\frac{1}{\left\langle b\right\rangle}\approx \frac{1}{6}\;$GeV$^2$.

\begin{figure}[h!]
\hspace*{\fill}
\begin{tabular}[p]{c}
\subfigure[AS (purple dashed line), WW (blue long dashed line) and Total (red solid line) contributions vs H1 data.]{\epsfxsize=12 cm\epsfbox{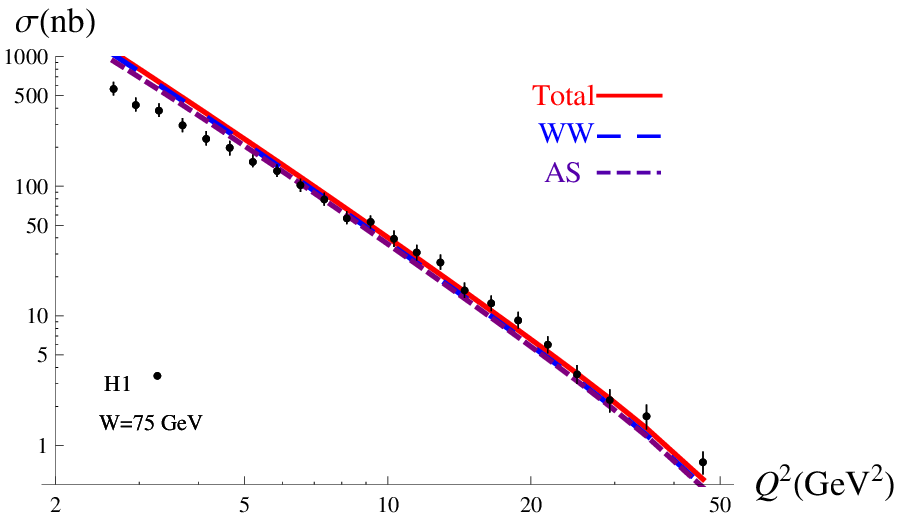}\label{sigH1}}\\
\subfigure[AS (purple dashed line), WW (blue long dashed line) and Total (red solid line) contributions vs ZEUS data.]{\epsfxsize=12 cm\epsfbox{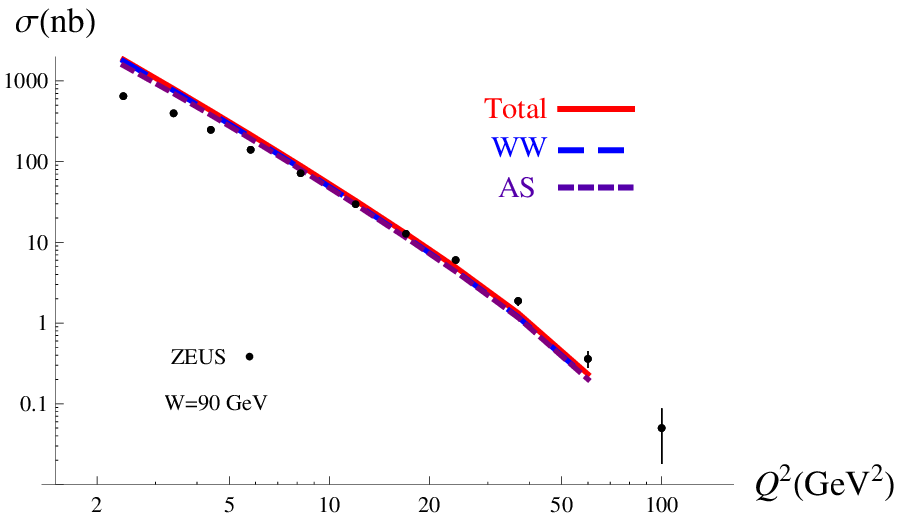}\label{sigZEUS}}
\end{tabular}
\hspace*{\fill}
\caption{Predictions for $\sigma$ vs $Q^2$ compared respectively with H1\cite{Aaron:2009xp} data (figure (a)) for $W=75\;$GeV and with ZEUS\cite{Chekanov:2007zr} data (figure (b)) for $W=90\;$GeV, using the AAMQSa-model.}
\label{FigAAa2}
\end{figure}

\begin{figure}[h!]
\hspace*{\fill}
\begin{tabular}[p]{c}
\subfigure[Asymptotic (purple dashed line), WW (blue long dashed line), total (red solid line) contributions vs H1 data.]{\epsfxsize=12 cm\epsfbox{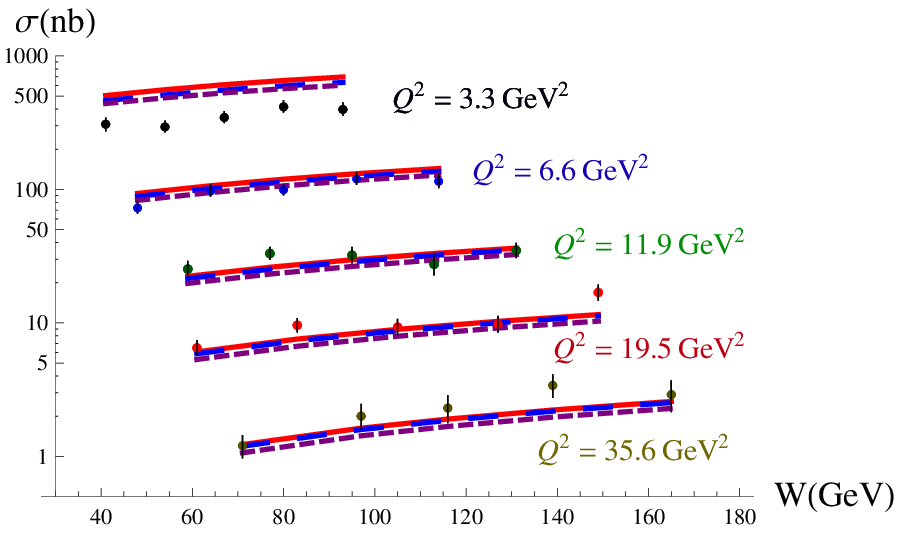}\label{sigWH1}}\\
 \subfigure[Total contribution to $\sigma$ including the $b-$slope errors vs H1 data.]{\epsfxsize=12 cm \epsfbox{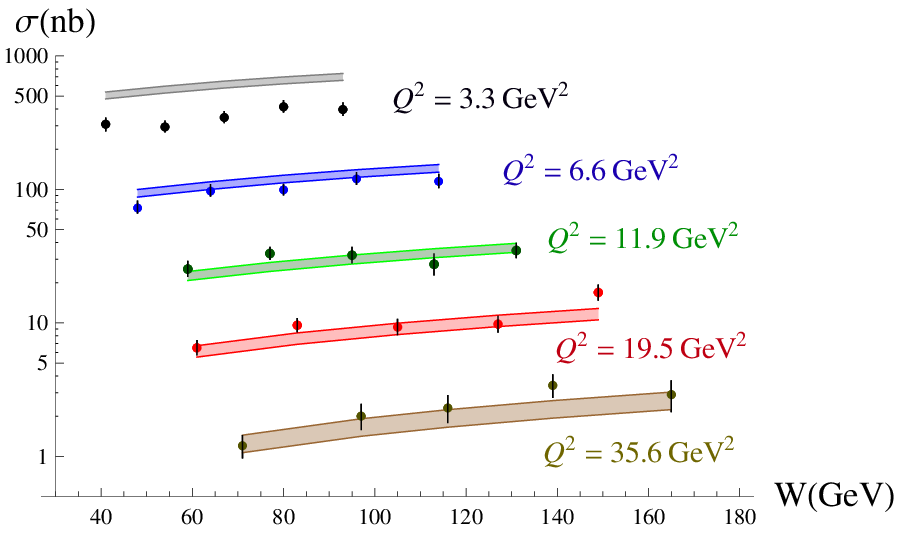}\label{sigbH1}}
\end{tabular}
\hspace*{\fill}
\caption{Predictions for $\sigma$ vs $W$, using the AAMQSa-model, compared with H1\cite{Aaron:2009xp} data. Figure (a): AS, WW and Total contributions. Figure (b): Total contribution taking into account the uncertainties on the $b-$slope. }
\label{FigAAa3}
\end{figure}

\begin{figure}[h!]
\hspace*{\fill}
\begin{tabular}[p]{c}
\subfigure[Asymptotic (purple dashed line), WW (blue long dashed line), total (red solid line) contributions vs ZEUS data.]{\epsfxsize=12 cm\epsfbox{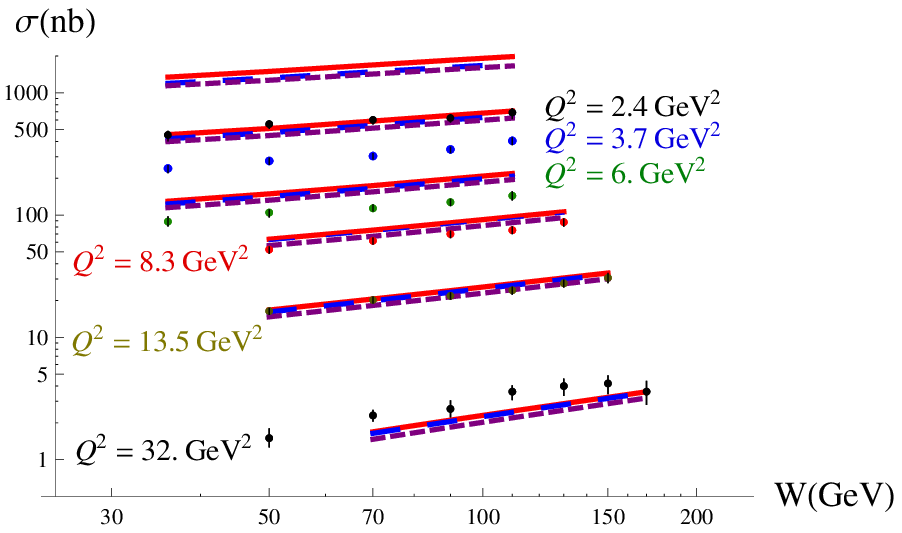}\label{sigWZEUS}}\\
\subfigure[Total contribution to $\sigma$ including the $b-$slope errors vs ZEUS data.]{\epsfxsize=12 cm\epsfbox{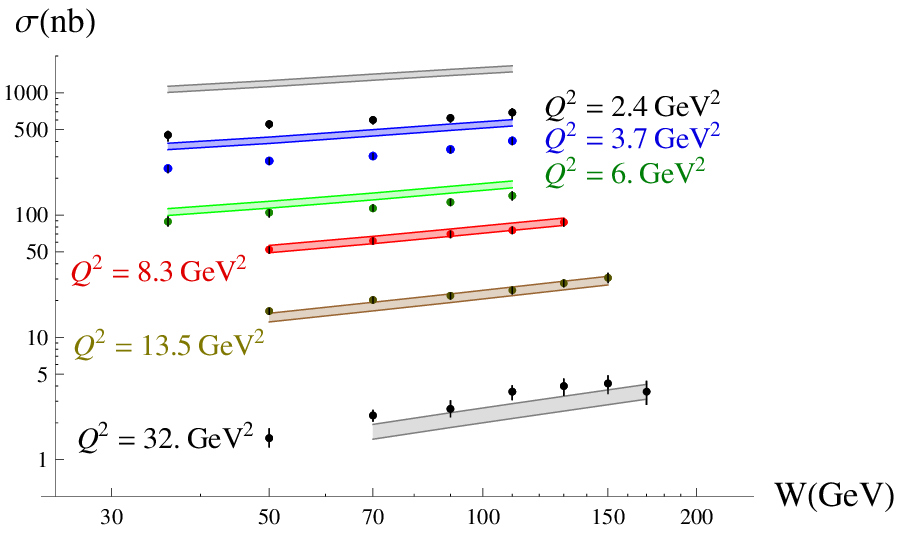}\label{sigbZEUS}}
\end{tabular}
\hspace*{\fill}
\caption{Predictions for $\sigma$ vs $W$, using the AAMQSa-model, compared with ZEUS\cite{Chekanov:2007zr} data. Figure (a): AS, WW and Total contributions. Figure (b): Total contribution taking into account the uncertainties on the $b-$slope. }
\label{FigAAa4}
\end{figure}

We now compare our predictions with the data for the total cross-section $\sigma$, given by the sum $\sigma=\sigma_L+\sigma_T$ according to ZEUS convention in ref.~\cite{Chekanov:2007zr} or $\sigma=\varepsilon \sigma_L+\sigma_T$ following H1 notation \cite{Aaron:2009xp}, where $\varepsilon$ is the photon polarization parameter\footnote{For H1 $\langle \varepsilon \rangle
=0.98$ and for ZEUS $\langle \varepsilon \rangle=0.996.$} 
\beq
\label{epsilon}
\varepsilon \simeq (1-y)/(1-y+y^2/2)\,.
\eq
We show in figures~\ref{sigH1} and  \ref{sigZEUS}, the AS, WW and Total contributions to the total cross-section $\sigma$ as function of $Q^2$ for fixed averaged $W$. The predictions are larger than the data for $Q^2$ smaller than approximately 7\;GeV$^2$, as expected from the results of $\sigma_{L,T}$. 
The figures \ref{sigWH1}, \ref{sigbH1}, \ref{sigWZEUS} and \ref{sigbZEUS} show the $W$ dependence of the cross-section for several values of $Q^2$. We see again 
 a good agreement between the predictions and the data for $Q^2$ approximately above 6\;GeV$^2$ for H1 data and 8\;GeV$^2$ for ZEUS data, taking into account the uncertainty on the $b-$slope.

\begin{figure}[h!]
\hspace*{\fill}
\begin{tabular}[p]{cc}
\hspace{-.3cm}\subfigure[AAMQSa versus H1 data.]{\epsfxsize=7.5 cm\epsfbox{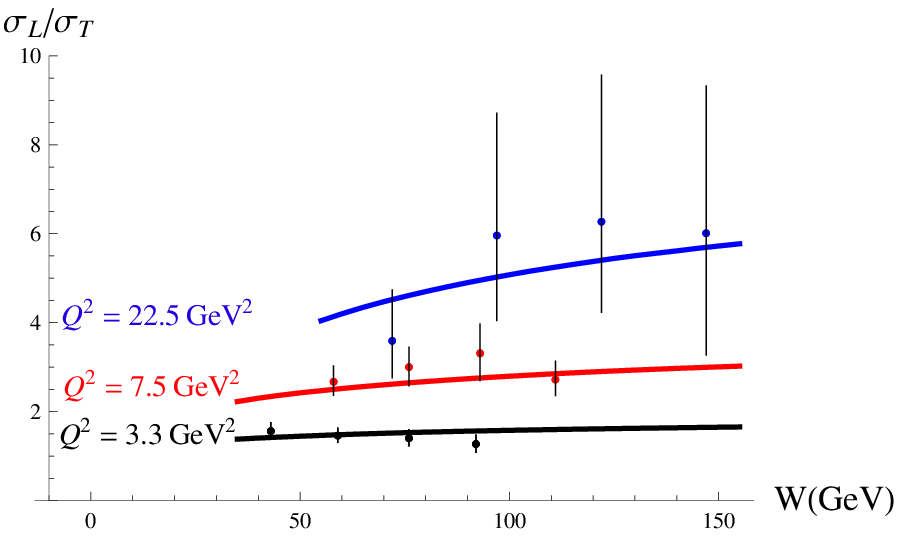}}
& \hspace{-.3cm}\subfigure[AAMQSa versus ZEUS data.] {\epsfxsize=7.5 cm \epsfbox{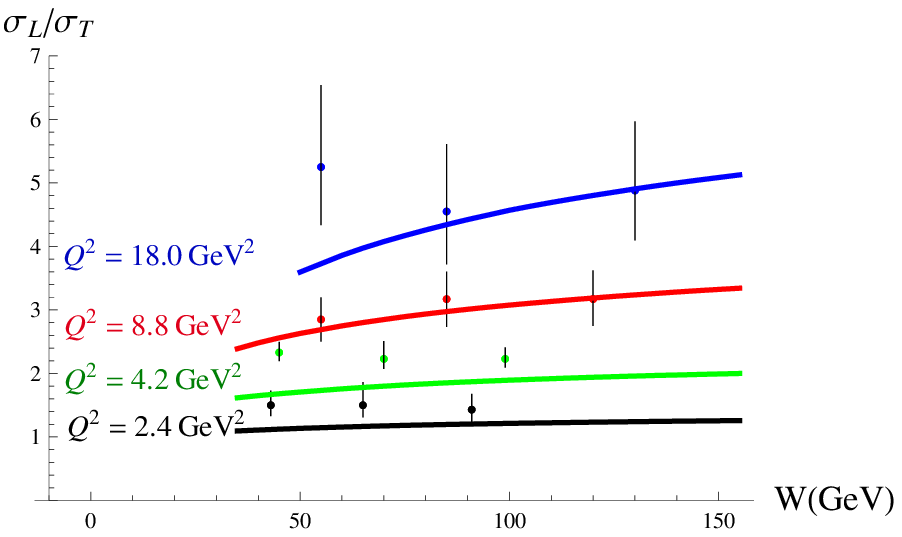}}
\end{tabular}
\hspace*{\fill}
\caption{The full twist~3 contribution to the ratio of the cross-sections $R=\sigma_L/\sigma_T$ in the limit $t=0$ versus $W$ and $Q^2$ compared to the data of H1 \cite{Aaron:2009xp} in figure (a) and ZEUS \cite{Chekanov:2007zr} in figure (b).}
\label{fig:AAaRatio}
\end{figure}

\begin{figure}[h!]
\hspace*{\fill}
\begin{tabular}[p]{c}
\subfigure[Asymptotic (purple dashed line), WW (blue long dashed line), Total (red solid line) contributions using AAMQSa-model vs H1 data.]{\epsfxsize=9 cm\epsfbox{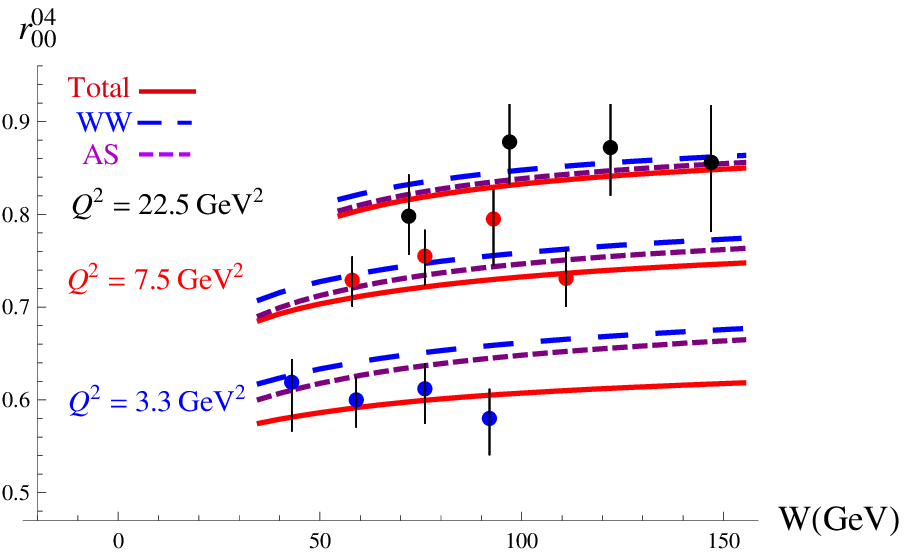}}\\
 \subfigure[Asymptotic (purple dashed line), WW (blue long dashed line), Total (red solid line) contributions using AAMQSa-model vs ZEUS data.]{\epsfxsize=9 cm \epsfbox{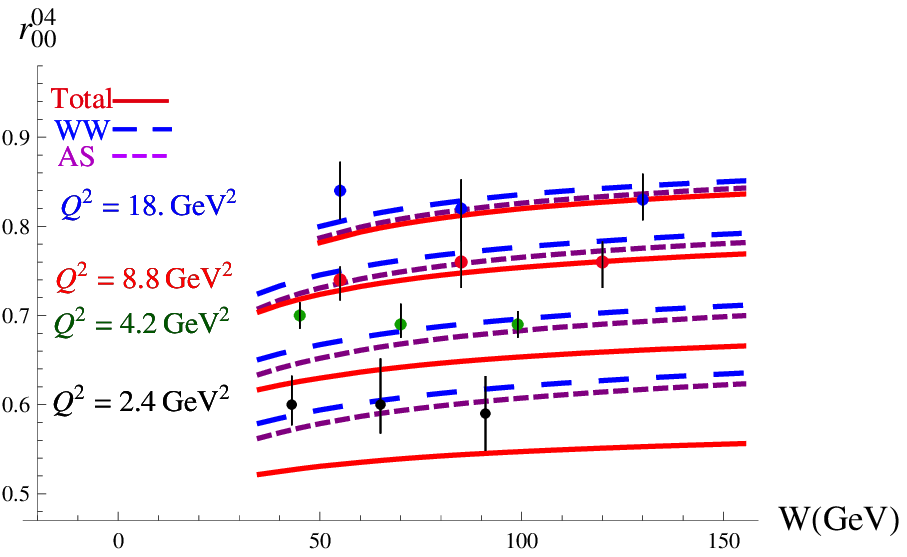}}
\end{tabular}
\hspace*{\fill}
\caption{Predictions for $r^{04}_{00}$ vs $W$ and $Q^2$ compared respectively with H1\cite{Aaron:2009xp} data (figures (a)) and ZEUS\cite{Chekanov:2007zr} data (figures (b)), using the AAMQSa-model.}
\label{Figr04}
\end{figure}

Finally, our analysis also provides predictions for the ratios $R$ and for the spin density matrix element $r^{04}_{00}$ 
\bea
\label{R-ratio}
R&=&\frac{\sigma_L}{\sigma_T} 
\,,\\
r^{04}_{00}&=&\frac{\sigma_L}{\sigma}\,.
\label{r04SCHC}
\eea
Assuming to keep only the SCHC amplitudes $T_{11}$ and $T_{00}$, and the equality of slopes $b_L=b_T$, the $t-$dependences of the cross-sections cancel in the ratios $R$ and $r^{04}_{00}$, leading to
\bea
\label{R-approx}
R &=& \frac{1}{x_{11}^2}
\eqa
and
\beqa
\label{r0_04_ratioT}
r^{04}_{00} &=&\frac{\varepsilon}{\varepsilon+x_{11}^2}\,,
\eea
where $x_{11}=|T_{11}|/|T_{00}|\,.$
H1 and ZEUS measurements of $R$ and $r^{04}_{00}=\sigma_L/\sigma$ as functions of $\left|t\right|$ confirm this weak dependence on $\left|t\right|$. Based on H1 data, we can estimate \cite{Anikin:2011sa} the correction to the ratio $r^{04}_{00}$  due to the amplitude $T_{01},$ for the $t-$range of H1, to be below 1$\%$.
The results are shown in figure~\ref{fig:AAaRatio} for the ratio $R$ and in figure~\ref{Figr04} for the spin density matrix element $r^{04}_{00}$,  using AAMQSa model.

\section{The radial distributions of dipoles involved in the overlap of the $\gamma^*_{L(T)}$ and  $\rho_{L(T)}-$meson states}
\label{sec:distribution}
In the dipole picture, 
the overlap of the wave functions of the outgoing $\rho$ meson $\Psi_{(h,\bar h)}^{*\rho_{L,T}\,(\lambda_{\rho})}$ and of the incoming virtual photon $\Psi_{(h,\bar h)}^{\gamma^*_{L,T}\,(\lambda_{\gamma})}$ 
 represents the amplitude of probability for these states to dissociate into a quark anti-quark color dipole 
 of size $\rb$, the quark having a longitudinal momentum fraction $y$. 
 We define thus the probability amplitude $\mathcal{W}_{\lambda_{\rho}\lambda_{\gamma}}$ as the corresponding parts of the impact factors appearing in eqs.~(\ref{T00Lpsi}) and (\ref{T11Tpsi}), 
\beq
\mathcal{W}_{00}(y,r;\mu_F^2,Q^2)=\psi^{\gamma^*_L\to\rho_L}_{(q\qb)}(y,\rb;Q,\mu_F^2)\,,
\eq
for the amplitude $T_{00}$ and
\bea
\label{W11}
\mathcal{W}_{11}(y,r;\mu_F^2,Q^2)&=&\psi^{\gamma^*_T\to\rho_T}_{(q\qb)}(y,\rb;Q,\mu_F^2)+\int_{0}^{y} d y_1\, \psi^{\gamma^*_T\to\rho_T}_{(q\qb g)}(y_1,y,\rb;Q,\mu_F^2)\,,\\
\label{W11WW}
\mathcal{W}^{WW}_{11}(y,r;\mu_F^2,Q^2)&=&\psi^{\gamma^*_T\to\rho_T\,WW}_{(q\qb)}(y,\rb;Q,\mu_F^2)\,,\\
\label{W11Gen}
\mathcal{W}^{gen}_{11}(y,r;\mu_F^2,Q^2)&=&\psi^{\gamma^*_T\to\rho_T\,\text{gen}}_{(q\qb)}(y,\rb;Q,\mu_F^2)+\int_{0}^{y} \,d y_1 \psi^{\gamma^*_T\to\rho_T}_{(q\qb g)}(y_1,y,\rb;Q,\mu_F^2)
\eea
for the Total, the WW and the genuine contributions to $T_{11}$. The probability amplitudes $\mathcal{W}_{\lambda_{\rho}\lambda_{\gamma}}$ permit in turn to define the radial distributions $\mathcal{P}_{\lambda_{\rho}\lambda_{\gamma}}$ of the interacting dipole, as
\bea
\label{P00}
\mathcal{P}_{\lambda_{\rho}\lambda_{\gamma}}(r,Q^2,\mu_F^2)&=& \frac{1}{\mathcal{N}_{\lambda_{\rho}\lambda_{\gamma}}}\,\displaystyle r \int dy \, \left| \mathcal{W}_{\lambda_{\rho}\lambda_{\gamma}}(y,r;\mu_F^2,Q^2) \right| \,,
\eea
 where $\mathcal{N}_{\lambda_{\rho}\lambda_{\gamma}}$ are normalization factors
\bea
\mathcal{N}_{\lambda_{\rho}\lambda_{\gamma}}&=&\displaystyle \int_0^{\infty}dr\,r  \int dy\,\left| \mathcal{W}_{\lambda_{\rho}\lambda_{\gamma}}(y_1,y,r;\mu_F^2,Q^2) \right|\,.
\eea
Expressed in terms of these functions, the scattering amplitudes read
\bea
\label{defImpep00P}
\frac{T_{\lambda_{\rho}\lambda_{\gamma}}}{s}&=&\mathcal{N}_{\lambda_{\rho}\lambda_{\gamma}}\,\int_{0}^{\infty} dr\, \mathcal{P}_{\lambda_{\rho}\lambda_{\gamma}}(r,Q^2,\mu_F^2) \, \hat{\sigma}(x,r)\,.
\eea
 The probability amplitude for a dipole of size $r$ to scatter on the nucleon is then proportional to  $\mathcal{P}_{\lambda_{\rho}\lambda_{\gamma}}(r,Q^2,\mu_F^2) \, \hat{\sigma}(x,r)$, which justifies the inclusion of the factor $r$ in (\ref{P00}).

Below we will also use the rescaled radial distributions $P_{\lambda_{\rho}\lambda_{\gamma}}(\lambda,\mu_F^2)$, 
\beq
\label{Plambda}
P_{\lambda_{\rho}\lambda_{\gamma}}(\lambda,\mu_F^2)\equiv \frac{\mathcal{P}_{\lambda_{\rho}\lambda_{\gamma}}(\frac{\lambda}{Q},Q^2;\mu_F^2)}{Q}\,,
\eq
 which depend on $r$ and $Q$ only through the variable $\lambda=r\, Q$ and we choose to put $\mu_F^2=\mu_F^2(Q^2)$, see eq.~(\ref{muF}). Note that the rescaled asymptotic distributions, 
\beq
\label{ASPlambda}
P_{\lambda_{\rho}\lambda_{\gamma}}^{(AS)}(\lambda)\equiv P_{\lambda_{\rho}\lambda_{\gamma}}^{(AS)}(\lambda,\mu_F^2=\infty)\,,
\eq
 are independent of $Q^2$. 
 
 This change of variable leads to the formulas
\bea
\label{defImpep00Plambda}
\frac{T_{\lambda_{\rho}\lambda_{\gamma}}}{s}&=&\mathcal{N}_{\lambda_{\rho}\lambda_{\gamma}}\,\int_{0}^{\infty} d\lambda\, P_{\lambda_{\rho}\lambda_{\gamma}}(\lambda,\mu_F^2) \, \tilde{\sigma}(x,\lambda)\,,
\eea
with 
\beq
\label{tildeSigma}
\tilde{\sigma}(x,\lambda)=\hat{\sigma}\left(x,\frac{\lambda}{Q}\right)\,.
\eq

The average value of a function $f(y)$ depending of the longitudinal fraction of momentum $y$ carried by one of the partons, will be estimated by  
\bea
\label{yy00}
\left\langle f(y)\right\rangle_{\lambda_{\rho}\lambda_{\gamma}}&=& \frac{1}{\mathcal{N}_{\lambda_{\rho}\lambda_{\gamma}}}\,\displaystyle \int dr \,\int dy \,f(y) \,r\, \left| \mathcal{W}_{\lambda_{\rho}\lambda_{\gamma}}(y,r;\mu_F^2,Q^2) \right| \,.
\eea

\subsection{The radial distribution of the $\gamma^*_L \to \rho_L$ transition}
\label{subsec:P00}

\begin{figure}[h!]
\hspace*{\fill}
\begin{tabular}[p]{cc}
\subfigure[Total $\mathcal{P}_{00}(r,Q^2,\mu_F^2(Q^2))$ (red solid), $\mathcal{P}^{(AS)}_{00}(r,Q^2)$ (red dashed) distributions and  $\hat{\sigma}(x,r)$ at $Q^2=1\;$GeV$^2$.] {\epsfxsize=7 cm \epsfbox{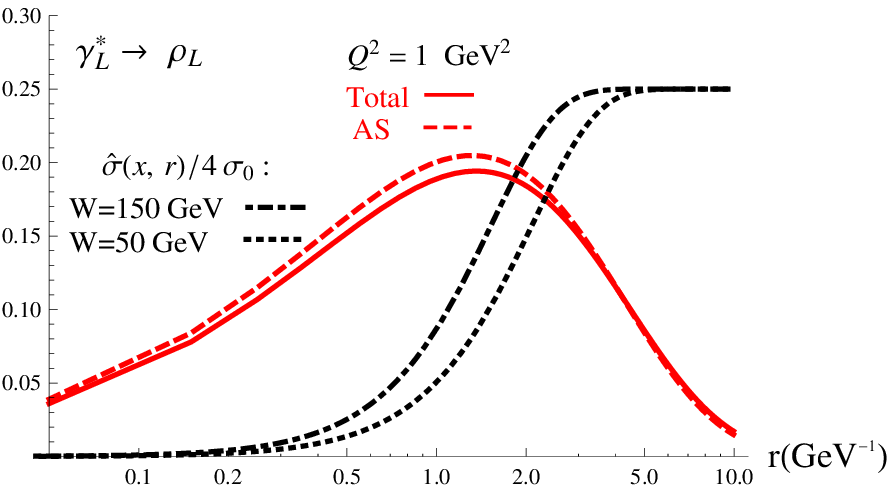}\label{ConvolL-1GeV}}& \subfigure[Total $\mathcal{P}_{00}(r,Q^2,\mu_F^2(Q^2))$ (red solid), $\mathcal{P}^{(AS)}_{00}(r,Q^2)$ (red dashed) distributions and  $\hat{\sigma}(x,r)$ at $Q^2=10\;$GeV$^2$.] {\epsfxsize=7 cm \epsfbox{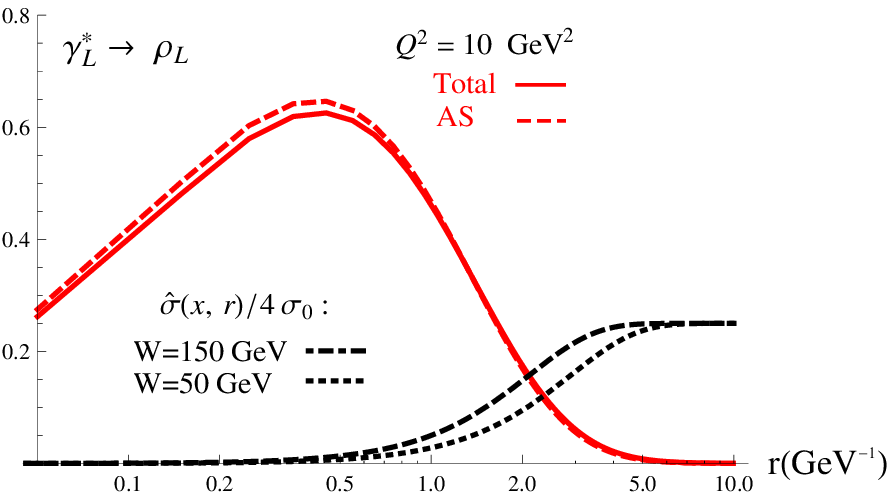}\label{ConvolL-10GeV}}
\end{tabular}
\hspace*{\fill}
\caption{The functions $\mathcal{P}_{00}(r,Q^2,\mu_F^2)$ and $\mathcal{P}^{(AS)}_{00}(r,Q^2)$ vs the size $r$ of the interacting dipole, for $Q^2=1\;$GeV$^2$ (\ref{ConvolL-1GeV}) and $Q^2=10\;$GeV$^2$ (\ref{ConvolL-10GeV}), and the dipole cross-section $\hat{\sigma}(x,r)$ normalized by the factor $4\sigma_0$ for $W=50\;$GeV and $W=150\;$GeV.}	
\label{fig:Fig-Convol}
\end{figure} 

The distributions $\mathcal{P}_{00}(r,Q^2,\mu_F^2)$ and $\mathcal{P}^{(AS)}_{00}(r,Q^2)\equiv \mathcal{P}_{00}(r,Q^2,\infty)$ are close to each other, as it is shown in figures~\ref{ConvolL-1GeV} and \ref{ConvolL-10GeV}, which indicates that the distribution $\mathcal{P}_{00}(r,Q^2,\mu_F^2)$ is not sensitive to $\mu_F^2$. We can then restrict the study of $\mathcal{P}_{00}(r,Q^2,\mu_F^2)$ by considering only $\mathcal{P}_{00}^{(AS)}(r,Q^2)$, which is also simpler for the analytic treatment. 
At first glance we see that the distributions are peaked around $r\sim \frac{1.3}Q$ and consequently the peak moves to the right and the distribution becomes wider as $Q^2$ decreases. Note that the dependency of $\hat{\sigma}(x,r)$ with respect to $Q^2$, which can be seen in figure~\ref{fig:Fig-Convol}, only occurs through the dependency of $R_0(x)$, according to eq.~(\ref{R0}).

\begin{figure}[htbp]
	\centering
		\includegraphics{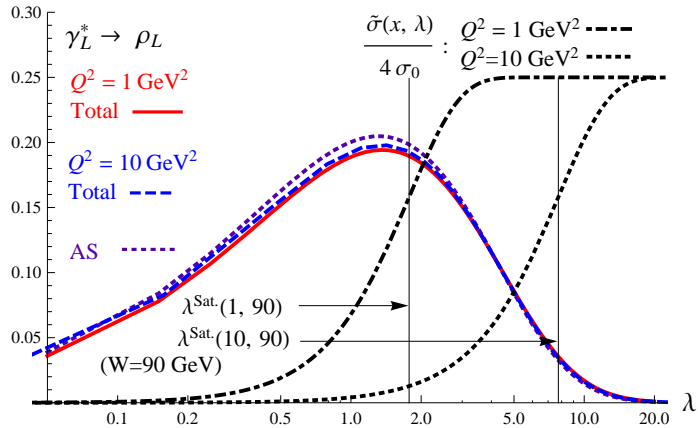}
\caption{Full twist~3 (Total) $P_{00}(\lambda,\mu_F^2(Q^2))$  for $Q^2=1$GeV$^2$ (solid red) and $Q^2=10$GeV$^2$ (dashed blue), AS $P^{(AS)}_{00}(\lambda)$ (dotted purple) and $\tilde{\sigma}(x,\lambda)$ at $W=90\;$GeV$^2$ for $Q^2=1\;$GeV$^2$(dotted-dashed black) and $Q^2=10\;$GeV$^2$ (dashed black).}
\label{Ptilde-L}
\end{figure}

In figure~\ref{Ptilde-L} we show the Total and AS rescaled radial distributions $P_{00}(\lambda,\mu_F^2(Q^2))$ and  $P_{00}^{(AS)}(\lambda,\mu_F^2(Q^2))$.  This last one 
 reads
\beq
\label{PLAS}
P_{00}^{(AS)}(\lambda)=\frac{1}Q \mathcal{P}_{00}^{(AS)}(\frac{\lambda}Q,Q^2)= 6 \,\int dy \,(y\yb)^2 \lambda\, K_0(\sqrt{y\yb} \lambda)\,.
\eq
The average value of $\lambda$ estimated with $P_{00}^{(AS)}(\lambda)$ is 
\beq
\label{lambdaMeanL}
\left\langle \lambda\right\rangle_{00}^{(AS)}=\int d\lambda \,\lambda \,P_{00}^{(AS)}(\lambda)= \frac{3\pi^2}{8}\approx 3.7\,.
\eq
About half of the dipoles are contained in the region $1<\lambda< \left\langle \lambda\right\rangle_{00}^{(AS)}\,,$ the peak of the distribution being at $\lambda^{\text{peak}}\sim 1.3$. The typical transverse scale $\mu=\sqrt{y\yb \, Q^2}$ entering the wave functions overlap  can be estimated using eq.~(\ref{yy00}), 
\beq
\label{muQ00AS}
\frac{\left\langle \mu\right\rangle_{00}^{(AS)}}{Q}=\left\langle \sqrt{y\yb}\right\rangle^{(AS)}_{00}=6\,\int dy \,\int d\lambda \,(\sqrt{y\yb})\,\left((y\yb)^{2} \lambda \, K_0(\sqrt{y\yb} \lambda)\right)=\frac{9\pi}{64}\approx 0.44\,.
\eq
The choice of the factorization scale $\mu_F(Q^2)$ given by eq.~(\ref{muF}) is then a good approximation of the transverse dynamical scale $\left\langle \mu\right\rangle^{(AS)}_{00}$ involved in the process.

The dipole scattering amplitude plays the role of a filter that selects dipoles of $\lambda > \lambda^{\text{Sat.}}(Q^2,W)=2 R_{0}(x)\,Q$. Note that the critical saturation line eq.~(\ref{Qsat}) is given 
 by
\beq
\label{criticalLine}
\lambda^{\text{Sat.}}(Q_S^2(x(Q_S^2,W^2)),W)=2\,,
\eq
where
\beq
\label{Eqsat}
x(Q^2,W^2)= \frac{Q^2}{W^2+Q^2}\left(1+\frac{4\, m_f^2}{Q^2}\right)\,,
\eq
in accordance with eq.~\ref{redefxBj}.
 In the kinematics of HERA, the energy in the center of mass $W$ varies roughly from $50\;$GeV to $150\;$GeV, leading to the following bounds for the two values $Q^2=1\;$GeV$^2$ and $Q^2=10\;$GeV$^2$,
\bea
\lambda^{\text{Sat.}}(1,50)&=&2.1 >\lambda^{\text{Sat.}}(1,W)>\lambda^{\text{Sat.}}(1,150)=1.5\\
\lambda^{\text{Sat.}}(10,50)&=&9.2 >\lambda^{\text{Sat.}}(10,W)>\lambda^{\text{Sat.}}(10,150)=6.7\,.
\eea
We will fix for our purpose $W=90\;$GeV resulting in the values, ($\lambda^{\text{Sat.}}(1\;\text{GeV}^2, 90\;$GeV)$\sim 1.8$) and ($\lambda^{\text{Sat.}}(10\;\text{GeV}^2, 90\;$GeV)$\sim 7.7$). We can then differentiate the case $Q^2=1\;$GeV$^2$ where we are in the saturation regime ($\lambda^{\text{Sat.}}(1\;\text{GeV}^2, 90\;$GeV)$<2$), and the case $Q^2=10\;$GeV$^2$ where saturation effect are less important ($\lambda^{\text{Sat.}}(10\;\text{GeV}^2, 90\;$GeV)$>2$).

\begin{figure}[htbp]
\begin{center}
	\includegraphics[width=0.65\textwidth]{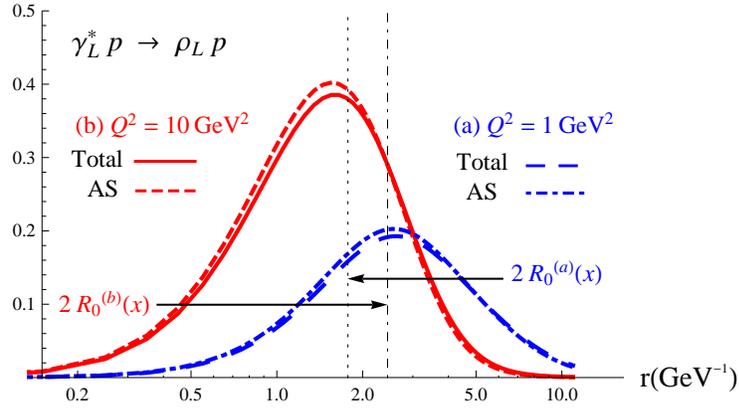}
\end{center}
	\caption{The normalized integrand of $T_{00}$, i.e. $\mathcal{P}_{00}(r,Q^2,\mu_F^2) \, \hat{\sigma}(x,r)$. The Total integrand at $\mu_F^2(Q^2)$ for $Q^2=1\;$GeV$^2$(blue long-dashed line) and $Q^2=10\;$GeV$^2$ (red solid line), and the AS integrand for $Q^2=1\;$GeV$^2$(blue dot-dashed line) and $Q^2=10\;$GeV$^2$ (red dashed line) integrands of $T_{00}$ for $W=90\;$GeV.}
\label{kGBW1100L}
\end{figure}
We can evaluate the percentages $N_{\lambda>\lambda^{\text{Sat.}}}$ of the dipoles large enough to be in the bandwidth of the dipole cross-section, 
 for each $Q^2$,
\bea
N_{\lambda > \lambda^{\text{Sat.}}}(Q^2=1\;\text{GeV}^2,\, W=90\,\text{GeV} )&=&\int_{\lambda^{\text{Sat.}}(1
,\,90
)}^{\infty} d\lambda \, P^{(AS)}_{00}(\lambda)=70\%\,, \\
N_{\lambda > \lambda^{\text{Sat.}}}(Q^2=10\;\text{GeV}^2, W=
90\,\text{GeV})&=&\int_{\lambda^{\text{Sat.}}(10
,\,90 
)}^{\infty} d\lambda\, P^{(AS)}_{00}(\lambda)=10\%\,,
\eea
as one can see in figure~\ref{Ptilde-L} (plotted in logarithmic scale).
The large difference between $N_{\lambda > \lambda^{\text{Sat.}}}(1,90)$ and $N_{\lambda > \lambda^{\text{Sat.}}}(10,90)$ indicates that the integrand of $T_{00}$ shown in figure~\ref{kGBW1100L}, is very sensitive, when $Q^2$ varies between 1\;GeV$^2$ and 10\;GeV$^2$, to the overlapping of the dipole cross-section bandwidth ($\lambda > \lambda^{\text{Sat.}}(Q^2,W)$) and the radial dipole distribution $P_{00}^{(AS)}(\lambda)$; we are then probing with a high accuracy the quality of the shape of the dipole cross-section.

At high $Q^2$ the tails of the distributions plays a dominant role. The tail of the distribution corresponds to the region where the integrand of the radial distribution can be approximated by an exponential fall,
\beq
\label{BesselK0tail}
\lambda \, K_{0}(\sqrt{y\yb}\lambda)\xrightarrow[]{\lambda \gtrsim \lambda^{\text{tail}}} \sqrt{\lambda}\, exp(-\sqrt{y\yb} \lambda)\,,
\eq
where typically
\beq
\lambda^{\text{tail}}\sim\frac{2}{\left\langle \sqrt{y\yb}\,\right\rangle_{00}^{(AS)}}\approx 4.5\,.
\eq

In the case $Q^2=1\;$GeV$^2$, the bandwidth of the dipole cross-section mostly overlaps with the peak of the distributions as 
$\lambda^{\text{Sat.}}(1
,90)\sim \lambda^{\text{peak}}$, while it only overlaps with the tail of the distribution when $Q^2=10\;$GeV$^2$, $ \lambda^{\text{Sat.}}(10,90)\sim \lambda^{\text{tail}}$.

Figure~\ref{kGBW1100L}
shows the normalized integrand of $T_{00}.$  This summarize our discussion on the respective role 
of the radial distribution and of the dipole cross-section.
This integrand is peaked near the saturation radius $r\sim 2R_0(x)$. Comparing the cases $Q^2=10\;$GeV$^2$ and $Q^2=1\;$GeV$^2$, we see that this peak is moving to the right as $Q^2$ decreases, going through the bandwidth of the dipole cross-section.

\subsection{The radial distribution of the $\gamma^*_T\to\rho_T$ transition}
\label{subsec:P11}
\begin{figure}[htbp]
	\centering
		\includegraphics[width=0.9\textwidth]{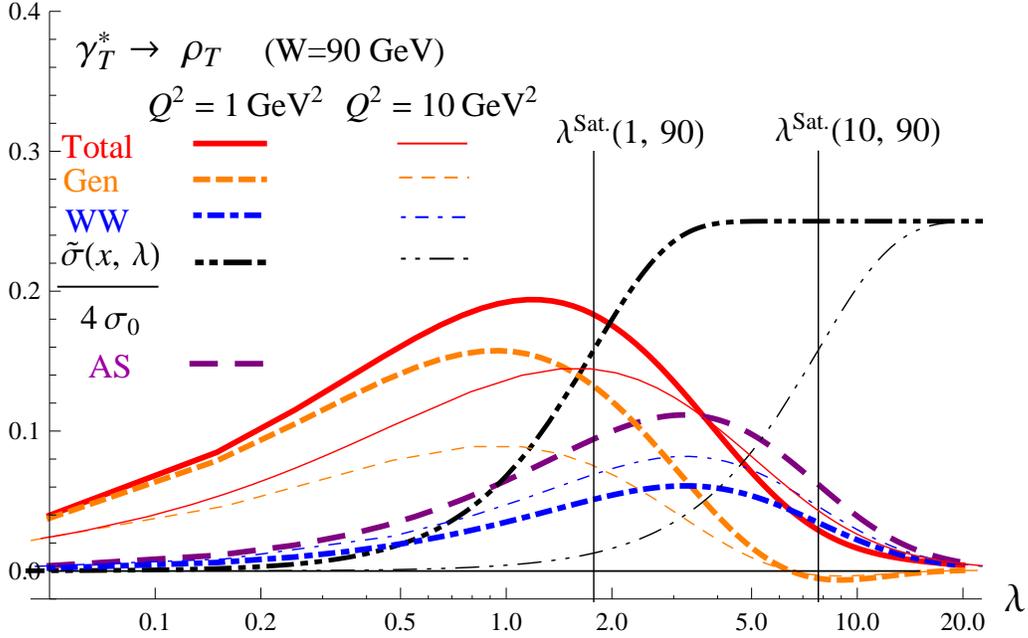}
\caption{The Total $P_{11}(\lambda,\mu_F^2(Q^2))$ (red solid lines) results and their WW (blue dot-dashed lines) and genuine (Gen) (orange dashed lines) contributions, as well as the AS (purple long-dashed line) result $P^{(AS)}_{11}(\lambda)$ and the dipole cross-sections $\tilde{\sigma}(x,\lambda)$ (black dot-dot-dashed lines) at $W=90\;$GeV$^2$, for $Q^2=1\;$GeV$^2$ (thick lines) and $Q^2=10\;$GeV$^2$ (thin lines).}
\label{PtildeTwwgen}
\end{figure}

In figure~\ref{PtildeTwwgen} 
are shown respectively for $Q^2=1\;$GeV$^2$ and $Q^2=10\;$GeV$^2$,
\begin{enumerate}
\item the full twist~3 (Total) rescaled radial distribution $P_{11}(\lambda,\mu_F^2(Q^2))$, where we distinguish the two following contributions,
\bei
\item the WW contribution $\tilde{P}^{WW}_{11}(\lambda,\mu_F^2(Q^2))$, 
\item the genuine contribution $\tilde{P}^{\text{gen}}_{11}(\lambda,\mu_F^2(Q^2))$,
\ei
such as\footnote{The tilde is to differentiate the contributions $\tilde{P}^{WW\,(\text{gen})}_{11}$ to the distribution $P_{11}$ from the distributions $P^{WW}_{11}$ or  $P^{(\text{gen})}_{11}$ which are normalized separately.} $P_{11}(\lambda,\mu_F^2)=\left|\tilde{P}^{WW}_{11}(\lambda,\mu_F^2)+\tilde{P}^{gen}_{11}(\lambda,\mu_F^2)\right|$,
\item the asymptotic rescaled radial distribution $P^{(AS)}_{11}(\lambda)$, using the asymptotic distribution amplitudes.
\end{enumerate}
Contrary to the $\gamma^*_L\to\rho_L$ transition, we see that the dependence on $\mu_F^2$ is quite strong if one compares $P_{11}(\lambda,\mu_F^2)$ to $P^{(AS)}_{11}(\lambda)$. We cannot then restrict ourselves to only study the asymptotic case.

It is interesting to estimate the average $\left\langle \lambda\right\rangle$ obtained with the different distributions $P_{11}$, $P^{(AS)}_{11}$, $P^{WW}_{11}$ and $P^{\text{gen}}_{11}$, where $P^{WW}_{11}$ and $P^{\text{gen}}_{11}$ have been normalized separately. The explicit expression for the asymptotic distribution
\beq
\label{PTAS}
P^{(AS)}_{11}(\lambda) =\frac{1}Q\mathcal{P}^{(AS)}_{11}\left(\frac{\lambda}Q,Q^2\right)=\frac{3}{4}\,\int dy \, (y\yb)^{3/2} (y^2+\yb^2)\,\lambda^2  K_1(\sqrt{y\yb}\lambda)\,,
\eq
leads to
\beq
\left\langle \lambda\right\rangle^{(AS)}_{11}=\int d\lambda \,\lambda \,P^{11, (AS)}(\lambda)= \frac{27\pi^2}{32}\approx 8.33\,.
\eq
The average value of $\lambda$ estimated with the WW distribution is, 
\bea
\left\langle\lambda\right\rangle^{WW}_{11}(\mu_F^2)&=&\int d\lambda \,\lambda\, P^{11,\,WW}(\lambda,\mu_F^2) \nn \\
&=&\frac{ \int d\lambda \,\lambda \int dy \,\left(y\yb \varphi_3^{WW}(y,\mu_F^2)\lambda K_1(\sqrt{y\yb}\lambda)\right)}{\int d\lambda \int dy \,\left(y\yb \varphi_3^{WW}(y,\mu_F^2)\lambda K_1(\sqrt{y\yb}\lambda)\right)}\nn\\
&=&\frac{9 \pi^2}{512}(48+13 a_2^{\|}(\mu_F^2))\xrightarrow[]{\mu_F^2\to \infty}\left\langle \lambda\right\rangle^{(AS)}_{11}\,.
\label{WWlambda}
\eea
The effect of the term involving $a_2^{\|}(\mu_F^2)$ in the r.h.s. of eq.~(\ref{WWlambda}) is under $4\%$, which indicates that $P^{WW}_{11}(\lambda,\mu_F^2)\sim P^{AS}_{11}(\lambda,\mu_F^2)$ is a good approximation. 
The computation of the average values of $\lambda$ for the all the different contributions are given in table~\ref{tab:fsq}.
\begin{table}[htbp]
	\centering
 \begin{tabular}{|l|c|c|c|r|}
 \hline
  & Total & WW& genuine& AS\\ \hline
$\left\langle \lambda\right\rangle_{11} (\mu_F^2(1\;\text{GeV}^2))$ & 6.3 & 8.7 & 3.2 & 8.3 \\ \hline
$\left\langle \lambda\right\rangle_{11} (\mu_F^2(10\;\text{GeV}^2))$ & 7.3 & 8.5 & 3.5 & 8.3 \\ \hline
\end{tabular}
\caption{Average values of $\left\langle \lambda\right\rangle=\left\langle r\,Q\right\rangle $ for the different contributions to the radial distribution for two values of  $\mu_F^2(Q^2)$.}
\label{tab:fsq}
\end{table}

The results in table~\ref{tab:fsq} show that the $\gamma^*_T\to\rho_T$ transition is more sensitive to saturation effects than the $\gamma^*_L\to\rho_L$ transition as $\left\langle \lambda\right\rangle_{11}$ is about twice larger than $\left\langle \lambda\right\rangle_{00}$. Indeed it means that more dipoles are produced in the bandwidth of the dipole cross-section 
 by the radial distribution $P_{11}(\lambda,\mu_F^2)$ than $P_{00}(\lambda,\mu_F^2)$. As a consequence the polarized cross-section $\sigma_T$ should be a more sensitive observable to 
probe features of the saturation regime than $\sigma_L$.

The transverse dipole scale associated to the WW contribution, using eq.~(\ref{yy00}), is 
\beq
\langle \mu \rangle^{WW}_{11}=Q\,\left\langle \sqrt{y\yb}\right\rangle_{11}^{WW} \approx \frac{Q}{2.7}\,. \nn
\eq
which is not so far from the values of the function $\mu_F^2(Q^2)=\frac{Q^2+M_{\rho}^2}{4}$ that is  used here.

The tail of the distribution $\tilde{P}^{WW}_{11}(\lambda,\mu_F^2)$ can be defined as,
\beq
\label{BesselK1tail}
\lambda^2 \, K_{1}(\sqrt{y\yb}\lambda)\xrightarrow[]{\lambda>\lambda^{\text{tail}\,WW}} \lambda \,\sqrt{\lambda} \, \exp(-\sqrt{y\yb} \lambda)\,,
\eq
where
\beq
\lambda^{\text{tail}\,WW}\sim \frac{4}{\left\langle \sqrt{y\yb}\right\rangle_{11}^{WW}}\approx 10.9 \,,
\eq
in which the average value $\left\langle \sqrt{y\yb}\right\rangle_{11}^{WW}$ is approximated by $\left\langle \sqrt{y\yb}\right\rangle_{11}^{AS}\approx 0.37$. 

 The genuine contribution $P^{\text{gen}}_{11}(\lambda,\mu_F^2)$, which vanishes in the limit $\mu_F^2\to \infty$, depends strongly on the factorization scale $\mu_F^2(Q^2)$, 
 as one can see in figure~\ref{PtildeTwwgen}. 
 The distribution $P^{\text{gen}}_{11}(\lambda,\mu_F^2)$ can be split into two contributions,
\beq
\label{Pgen}
 P^{\text{gen}}_{11}(\lambda,\mu_F^2)=\tilde{P}^{\text{gen}\,(q\qb)}_{11}(\lambda,\mu_F^2)+\tilde{P}^{\text{gen}\,(q\qb g)}_{11}(\lambda,\mu_F^2)\,,
\eq
where
\bea
\label{PqqbGEN}
\tilde{P}^{\text{gen}\,(q\qb)}_{11}(\lambda,\mu_F^2)& = & \frac{\lambda }{Q^2 {\cal N}_{11}} \int dy \, \psi^{\gamma^*_T\to\rho_T\,\text{gen}}_{(q\qb)}(y,\lambda/Q ; Q , \mu_F^2)\,,\\
\tilde{P}^{\text{gen}\,(q\qb g)}_{11}(\lambda,\mu_F^2) & = &  \frac{\lambda }{Q^2 {\cal N}_{11}}\int dy \int_{0}^{y} d y_1 \, \psi^{\gamma^*_T\to\rho_T}_{(q\qb g)}(y_1,y,\lambda/Q;Q,\mu_F^2)
\,.
\label{PqqbgGEN}
\eea
$\tilde{P}^{\text{gen}\,(q\qb)}_{11}(\lambda,\mu_F^2)$ involves the exchange of two-parton in the hard part and the genuine solutions of the two-parton  DAs, while $\mathcal{P}^{\text{gen}\,(q\qb g)}_{11}(\lambda,\mu_F^2)$ involves the three-parton exchange hard part $\mathcal{F}^{\gamma^*_T }(y_1,y_2,\rb;Q)$ defined in eq.~(\ref{F3part}).

The transverse scale $\mu$ present in $\tilde{P}^{\text{gen}\,(q\qb)}_{11}(\lambda,\mu_F^2)$  
 is of order $\left\langle \sqrt{y\yb}\right\rangle_{11}^{\text{gen}\,(q \qb)}(\mu_F^2)\sim 0.5$ and is not very sensitive to $\mu_F^2$. The choice of $\mu_F^2(Q^2)\sim \frac{Q}2$ is then a good choice for this contribution.

The contribution $\tilde{P}^{\text{gen}\,(q\qb g)}_{11}(\lambda,\mu_F^2)$ have several transverse scales which are $\mu_1$, $\mu_2$, $\mu_{\qb g}$, $\mu_{qg}$ and $\mu_{q\qb}$ defined by eqs.~(\ref{defmuGEN}), each corresponding to a dipole configuration involving two of the three partons available in the process. 
In order to estimate these transverse scales, we first evaluate the average fraction of momentum carried by the quark $\left\langle y_1\right\rangle_{11}^{\text{gen}\,(q \qb g)}(\mu_F^2)$, the antiquark  $\left\langle \yb_2\right\rangle_{11}^{\text{gen}\,(q \qb g)}(\mu_F^2)$ and the gluon $\left\langle y_g\right\rangle_{11}^{\text{gen}\,(q \qb g)}(\mu_F^2)$, defined as
\beq
\label{Average}
\left\langle y_i \right\rangle_{11}^{\text{gen}\,(q \qb g)}(\mu_F^2)= \frac{\displaystyle \int d\lambda\,\lambda\, \int dy_2 \,\int_{0}^{y_2} d y_1\,  y_i\, \psi^{\gamma^*_T\to\rho_T}_{(q\qb g)}(y_1,y_2,\rb;Q,\mu_F^2)}
{\displaystyle \int d\lambda\,\lambda\, \int dy_2 \,\int_{0}^{y_2} d y_1\,  \psi^{\gamma^*_T\to\rho_T}_{(q\qb g)}(y_1,y_2,\rb;Q,\mu_F^2)}\,.
\,
\eq
Using the fact that $\left\langle y_1 \right\rangle_{11}^{\text{gen}\,(q \qb g)}(\mu_F^2)=\left\langle \yb_2 \right\rangle_{11}^{\text{gen}\,(q \qb g)}(\mu_F^2)$ due to the symmetry under the exchange of the quark and the antiquark, the obtained values are given in table~\ref{Taby}. 
\begin{table}[htbp]
	\centering
 \begin{tabular}{|c|c|c|}
 \hline
    $\left\langle y_1 \right\rangle_{11}^{\text{gen}\,(q \qb g)}(\mu_F^2) $& $\left\langle \yb_2 \right\rangle_{11}^{\text{gen}\,(q \qb g)}(\mu_F^2) $& $\left\langle y_g \right\rangle_{11}^{\text{gen}\,(q \qb g)}(\mu_F^2) $\\ \hline
 41.8\% & 41.8\% & 16.3\%  \\ \hline
\end{tabular}
\caption{Average values  $\left\langle y_i \right\rangle_{11}^{\text{gen}\,(q \qb g)}(\mu_F^2) $ of the fraction of longitudinal momentum of the $\rho$ meson carried by each of the parton $i$.}
\label{Taby}
\end{table}

\noindent
An estimation of the transverse scales can be made, using 
\bea
\label{scales3parton1}
\frac{\left\langle \mu_1 \right\rangle}{Q} = \frac{\left\langle \mu_2 \right\rangle}{Q} & \sim & \sqrt{\left\langle y_1 \right\rangle (1-\left\langle y_1 \right\rangle)} \approx 0.49\,,\\
\label{scales3parton2}
\frac{\left\langle \mu_{q g} \right\rangle}{Q} = \frac{\left\langle \mu_{\qb g} \right\rangle}{Q} & \sim & \sqrt{ \frac{\left\langle y_1 \right\rangle\,\left\langle y_g \right\rangle}{(1-\left\langle \yb_2 \right\rangle)}}\approx 0.34\,,\\
\label{scales3parton3}
\frac{\left\langle \mu_{q \qb} \right\rangle}{Q} & \sim & \sqrt{\frac{ \left\langle y_1 \right\rangle\,\left\langle \yb_2 \right\rangle}{(1-\left\langle y_g \right\rangle)}}\approx 0.46\,,
\eea
where we assume 
that the transverse scale values are roughly approximated by using the average fractions of longitudinal momentum in eqs.~(\ref{defmuGEN}).

These values are evaluated at $\mu_F^2(1\;$GeV$^2)$. Other values of $\mu_F^2$ have also been used, leading to approximately the same results. We note that the function $\mu_F(Q^2)\approx \frac{Q}2$ is close to the $\left\langle \mu_1 \right\rangle$ and $\left\langle \mu_{q\qb} \right\rangle$ values, and of the same order of magnitude than $\left\langle \mu_{q g} \right\rangle$ but in principle, one should adapt the choice of the factorization scale to the relevant transverse scales at stake for each part of the process.

The WW and the genuine contributions to the radial distribution $P_{11}(\lambda,\mu_F^2)$ are of the same order of magnitude for $Q^2\sim 10\;$GeV$^2$, and the genuine contribution becomes even more important for $Q^2=1\;$GeV$^2$. 
At $Q^2=10\;$GeV$^2$, as
\beq
\left\langle \lambda\right\rangle_{11}^{\text{gen}}(\sim 3.5)\,<\,\lambda^{\text{Sat.}}(10\;\text{GeV}^2)(\sim 7.7)\,<\,\left\langle \lambda\right\rangle^{WW}_{11}(\sim 8.7)\,,\nn
\eq
 most of the dipoles in the bandwidth of the dipole cross-section are provided by the WW contribution, which explains why the predictions for $\sigma_T$ are dominated by the WW predictions, as shown in figure~\ref{sigT}, and consequently, why the results depend weakly on the factorization scale.

\begin{figure}[htbp]
\begin{center}
	\includegraphics[width=0.65\textwidth]{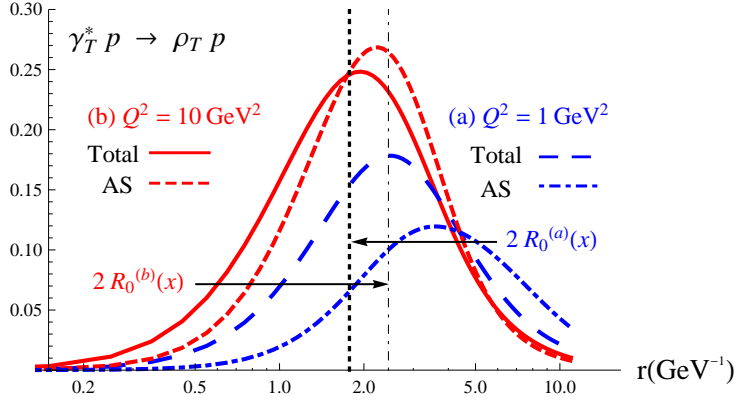}
\end{center}
	\caption{
	The Total contributions at $\mu_F^2(Q^2)$ for $Q^2=1\;$GeV$^2$ (blue long-dashed line) and $Q^2=10\;$GeV$^2$ (red solid line), and the AS contributions for $Q^2=1\;$GeV$^2$ (blue dot-dashed line) and $Q^2=10\;$GeV$^2$ (red dashed line) to the normalized integrands of $T_{11}$, i.e. $\mathcal{P}_{11}(r,Q^2,\mu_F^2) \hat{\sigma}(x,r)$, for $W=90\;$GeV.}
\label{kGBW1100T}
\end{figure}

 However the fact that the genuine contribution is important even at large $Q^2$, indicates that the three-parton exchange between the $\gamma^*_T$ and $\rho_T$ states is important in such relations as the normalization eq.~(\ref{Normalisation}) of the  $\rho$ meson wave function \cite{Lepage:1980fj, Munier:2001nr}, and the electronic decay width eq.~(\ref{decay}) \cite{Dosch:1996ss,Munier:2001nr},
 \bea
\label{Normalisation}
1&=&\sum_{h,\tilde{h}}\int dy \int d^2\rb \,\left| \Psi^{\rho_T}_{h,\tilde{h}}(y,\rb) \right|^2\,,\\
e f_{\rho} m_{\rho} (e^{*}_{\gamma}\cdot e_{\rho}) &=& \sum_{h,\tilde{h}}\int dy \int d^2\rb \, \Psi^{\rho_T}_{h,\tilde{h}}(y,\rb) \, \Psi^{\gamma^*_T}_{h,\tilde{h}}(y,\rb)\,,
\label{decay}
\eea
where the exchange of only two-parton is assumed. Indeed, the r.h.s. of eq.~(\ref{decay}), if one expands at large $Q^2$ the $\rho$ meson wave function
around $\rb=0\,,$
is the WW approximated result, which therefore misses the 
 genuine contributions arising from three-parton correlators, which can have a significant effect even for large $Q^2$ values, see figure~\ref{PtildeTwwgen}.
 These relations are usually 
used to constrain the parameters of the wave function models of the $\rho$ meson, assuming that the meson is solely constituted of a quark and an antiquark, which then consist in neglecting the higher Fock state contributions like the genuine contribution. 
At $Q^2=1\;$GeV$^2$, the peak of the genuine part of the radial distribution enters the dipole cross-section bandwidth ($\lambda^{\text{Sat.}}(1,90)<\left\langle \lambda\right\rangle^{\text{gen}}_{11}$) and gives an important contribution to the integrand of $T_{11}$. When increasing $Q^2$, as shown 
in figure~\ref{kGBW1100T} where we display the product $\mathcal{P}_{11}(r,Q^2,\mu_F^2) \hat{\sigma}(x,r)$ for $Q^2=1\;$GeV$^2$ and $Q^2=10\;$GeV$^2$, we can note that the difference between the AS and the Total results is a consequence of the genuine contribution growing when $Q^2$ decreases. It is the convolution with the dipole cross-section which washes-out the effect of these genuine twist-3 contributions.

\subsection{Comparison with the radial distributions obtained from models of the $\rho$ meson wave function.}
\label{subsec:rhoWaveFunction}

It is instructive to  compare shapes of the radial distributions $\mathcal{P}_{00}$ and $\mathcal{P}_{11}$ used in our analysis with those used in two other approaches which involve the overlap  of the virtual photon wave functions and of the $\rho$ meson wave functions:
\bei
\item  the "Boosted Gaussian" (BG) model \cite{Forshaw:2003ki},
\item  the "Gaus-LC" model \cite{Kowalski:2003hm}.
\ei
We use here the convention and the parameter values of ref.\cite{Kowalski:2006hc}, which for completeness are shown in table~\ref{TabModels}. 
\begin{table}[htbp]
	\centering
 \begin{tabular}{|c|c|c|c|c|c|c|c|}
 \hline
   Model & $N_T$ & $R_T^2\;$GeV$^{-2}$ & $N_L$ & $R_L^2\;$GeV$^{-2}$ & $f_{\rho}^{T}$\\ \hline
Gaus-LC &    4.47 & 21.9 & 1.79 & 10.4 & $f_{\rho}$\\ \hline 
  Boosted Gaussian   & 0.911 & 0.853 & 12.9 & $R_L^2$ & 0.182 \\ \hline
\end{tabular}
\caption{Parameter of the "Gaus-LC" and the "Boosted Gaussian" models taken from ref.\cite{Kowalski:2006hc}, for $M_{\rho}=0.776\;$GeV, $f_{\rho}=0.156\;$GeV, $m_f=0.14\;$GeV and with $f_{\rho}^{L}=f_{\rho}$ .
}
\label{TabModels}
\end{table}
The scalar parts of the wave functions are given by,
\bea
\label{GaussLCT}
\phi^{\text{Gauss-LC}}_{T}(y,r) &=& N_T \, (y\,\yb)^2\, e^{-\frac{r^2}{2R_T^2}}\,,\\
\label{GaussLCL}
\phi^{\text{Gauss-LC}}_{L}(y,r) &=& N_L \, y\,\yb \,e^{-\frac{r^2}{2R_L^2}}\,,\\
\label{BG}
\phi^{\text{BG}}_{L,T}(y,r)&=&N_{L,T}\, y\,\yb\, \exp\left(-\frac{m_f^2 R_{L,T}^2}{8y\yb}-\frac{2y\yb r^2}{R_{L,T}^2}+\frac{m_f^{2}R_{L,T}^2}{2}\right)\,.
\eea
The overlaps with the virtual photon wave function are, 
\bea
\label{OverlapT}
\sum_{h,\bar{h}}\Psi^{\rho_T*}_{h,\tilde{h}}(y,\rb) \Psi^{\gamma^*_T}_{h,\tilde{h}}(y,\rb)&\propto & m_f^2 K_0(\mu r) \phi_T(y,r)-(y^2+\yb^2) \mu K_1(\mu r)\partial_r \phi_T(y,r)\,,\\
\sum_{h,\bar{h}}\Psi^{\rho_L*}_{h,\tilde{h}}(y,\rb) \Psi^{\gamma^*_L}_{h,\tilde{h}}(y,\rb)&\propto & y\yb K_0(\mu r) \left(m_{\rho} \phi_L(y,r)+\delta \frac{m_f^2-\nabla_r^2}{m_{\rho}y\yb}\phi_L(y,r)\right)\,,
\label{OverlapL}
\eea
with $\delta=0$ for the Gaus-LC model and $\delta=1$ for the BG model. The radial distributions thus read, 
\beq
\label{Overlapdistrib}
\mathcal{P}_{L,T}(r)=\frac{1}{\mathcal{N}_{L,T}}r \int dy\,\sum_{h,\bar{h}}\Psi^{\rho_{L,T}*}_{h,\tilde{h}}(y,\rb) \Psi^{\gamma^*_{L,T}}_{h,\tilde{h}}(y,\rb)\,,
\eq
where the factors $\mathcal{N}_{L,T}$ normalize the distributions $\mathcal{P}_{L,T}(r)$.
\begin{figure}[h!]
\hspace*{\fill}
\begin{tabular}[p]{cc}
\subfigure[$\gamma^*_L\to\rho_L$ radial distributions and $\hat{\sigma}$ at $Q^2=1\;$GeV$^2$.] {\epsfxsize=7 cm \epsfbox{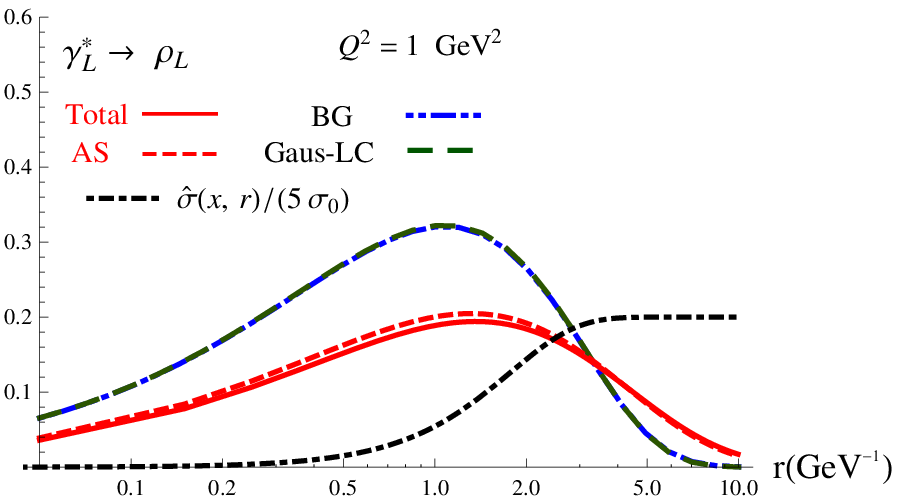}\label{OvFSSL1}}& \subfigure[$\gamma^*_L\to\rho_L$ radial distributions and $\hat{\sigma}$ at $Q^2=10\;$GeV$^2$.] {\epsfxsize=7 cm \epsfbox{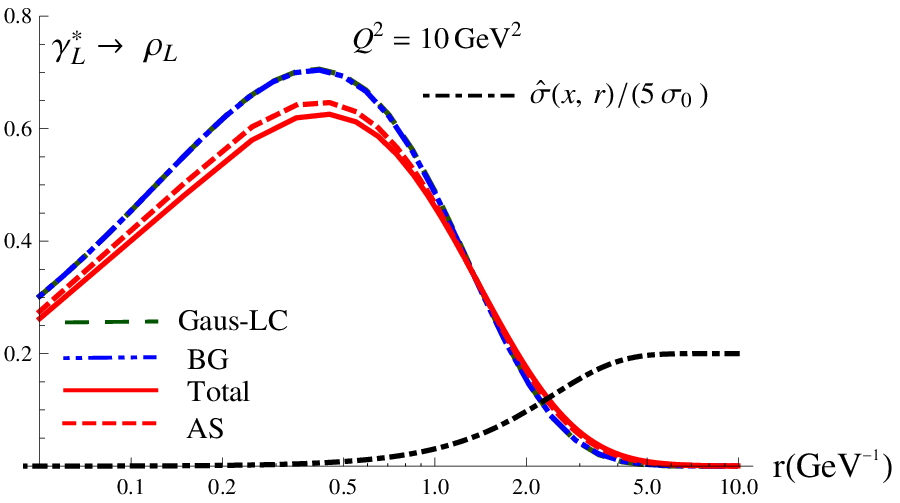}\label{OvFSSL10}}\\
\subfigure[$\gamma^*_T\to\rho_T$ radial distributions and $\hat{\sigma}$ at $Q^2=1\;$GeV$^2$.] {\epsfxsize=7 cm \epsfbox{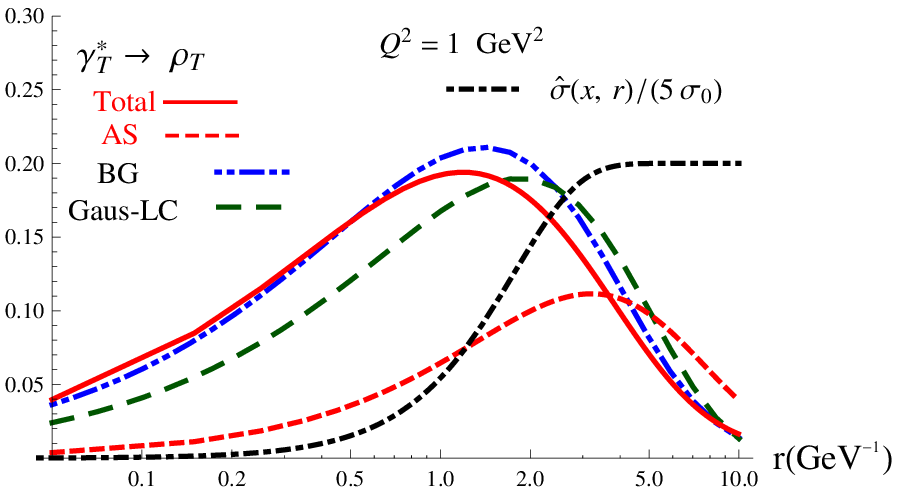}\label{OvFSST1}}& \subfigure[$\gamma^*_T\to\rho_T$ radial distributions and $\hat{\sigma}$ at $Q^2=10\;$GeV$^2$.] {\epsfxsize=7 cm \epsfbox{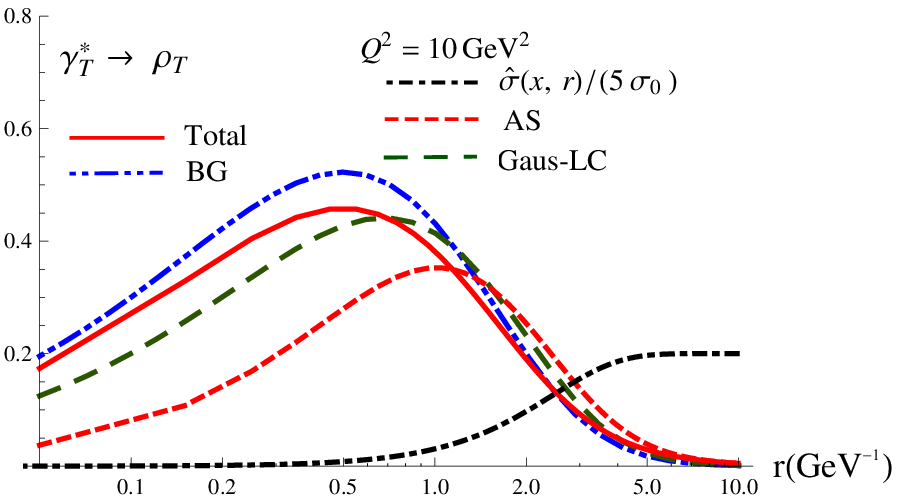}\label{OvFSST10}}
\end{tabular}
\hspace*{\fill}
\caption{The Gauss-LC (green, long dashed), BG (blue, dot-dot-dashed), Total (red, solid) and AS (red, dashed) radial distributions for the $\gamma^*_L\to\rho_L$ transition 
 (top) and for the $\gamma^*_T\to\rho_T$ transition 
  (bottom), vs $r$ for $Q^2=1\;$GeV$^2$ (left) and $Q^2=10\;$GeV$^2$ (right), as well as the dipole cross-section $\hat{\sigma}(x,r)$ rescaled by the factor $5 \sigma_0$ for $W=90\;$GeV (black, dot-dashed).}	
\label{OverlapFSS}
\end{figure}
Comparing the distributions at $Q^2=1\;$GeV$^2$ (figures~\ref{OvFSSL1} and \ref{OvFSST1}) and $Q^2=10\;$GeV$^2$ (figures~\ref{OvFSSL10} and \ref{OvFSST10}) we see that at large $Q^2$, in the bandwidth of the dipole cross-section, our distributions are converging with the distributions obtained from the Gaus-LC and BG models.

In the $\gamma^*_T\to\rho_T$ case in figures~\ref{OvFSST1} and~\ref{OvFSST10}, we see that the  BG and Gaus-LC models are closer to the distribution $\mathcal{P}_{11}$ than to the asymptotic distribution. When compared to other distributions, the asymptotic distribution $\mathcal{P}_{11}^{AS}$ is shifted to the right, thus  selecting larger dipole sizes. 

In the $\gamma^*_L\to\rho_L$ case, in figure~\ref{OvFSSL1} we see that the distributions from the Gaus-LC and BG models are not close to our predictions, indicating that the higher twist corrections are presumably more important at small $Q^2$  than in the $\gamma^*_T\to\rho_T$ transition.

These qualitative remarks remains the same when using the AAMQS dipole models, and we expect that they are model independent.


\section{Conclusions}
\label{sec:conclusion}
We performed phenomenological analysis of experimental data from HERA on $\rho$ meson electroproduction within the approach based on the recently derived~\cite{Besse:2012ia}  impact parameter representation of $\gamma^*\to\rho$ impact factor up to
twist-3 accuracy. The important feature of this representation consists  
in the inclusion of contributions coming from both  two- and three-partonic 
Fock states, maintaining a close connection with the  dipole model picture. Consequently, it was possible to include in our 
framework the saturation effects.        %
Our predictions show that we can get simultaneously good predictions for the polarized cross-sections $\sigma_{T}$ and $\sigma_L$.
The abilility of the model to reproduce the data is the confirmation of the following points:
\bei
\item the factorization of the dipole cross-section in the helicity amplitudes of the electroproduction of the $\rho$ meson works and, as an universal quantity, is the same for $T_{00}$ and $T_{11}$, giving the good energy dependence and normalizations of the polarized cross-sections,
\item the collinear factorization procedure of the $\rho$ meson is justified and works successfully beyond the leading twist.

\ei
 As expected the model has some limits due to the truncation of the twist expansion. Thanks to HERA data, we have identified 
  the virtuality $Q^{2 \, {\rm min}}\sim 5\;$GeV$^2$ were the higher twist corrections become important, which is a motivation to compute impact factors beyond the twist~3 accuracy in order to probe the genuine saturation regime which starts at $Q_S^2\sim~\!\!1\,$GeV$^2$.
 
  Other helicity amplitudes could be computed keeping the same approach, they would be useful in the $t\neq t_{\rm min}$ regime. The kinematics of the impact factor can be also extended to take into account the $t-$dependence of the impact factors, which would be a test for the dipole models which include the impact parameter dependence, 
 providing a probe of the proton shape \cite{Munier:2001nr}, in particular through local geometrical scaling \cite{Ferreiro:2002kv, Munier:2003bf}.

Data also exist for $\phi$ leptoproduction. In this case quark-mass effects should be taken into account, in particular, because this allows the transversely polarized $\phi$ to couple through its chiral-odd twist-2 DA.
Indeed, as it was pointed out in \cite{Anikin:2011sa}, the fact that the ratio $T_{11}/T_{00}$ is not the same (after trivial mass rescaling) for $\rho$ and $\phi$ mesons points to the importance of this effect.
This is also beyond the scope of our present study, but may open an interesting way for accessing chiral-odd DAs.

The
next-to-leading order effects - both on the evolution and on the impact factor - should be studied, since it is now known that both may have a important phenomenological effect \cite{Ivanov:2005gn, Ivanov:2006gt, Caporale:2007vs, Colferai:2010wu, Ducloue:2012bm}. 

On the experimental side, 
the future Electron-Ion Collider \cite{Boer:2011fh}
and Large Hadron Electron Collider \cite{AbelleiraFernandez:2012cc}
 with a high center-of-mass energy and high luminosities, as well as the International Linear Collider \cite{:2007sg, Djouadi:2007ik, Behnke:2007gj} will hopefully open  the opportunity  to study in more detail the hard diffractive production of mesons \cite{Pire:2005ic,Enberg:2005eq, Ivanov:2005gn,Pire:2006ik,Ivanov:2006gt,Segond:2007fj,Caporale:2007vs,Segond:2008ik}.

\vskip.2in \noindent
\acknowledgments

\noindent
We acknowledge Markus Diehl, Bertrand Duclou\'e, Krzysztof Golec-Biernat, Cyrille Marquet,  Leszek Motyka, St\'ephane Munier, Bernard Pire and Mariusz Sadzikowski for discussions and comments. Work supported by Polonium agreement.

This work is partly supported by the Polish Grant NCN
No. DEC-2011/01/B/ST2/03915, the French-Polish collaboration agreement 
Polonium
and the Joint Research Activity Study of Strongly 
Interacting Matter (acronym HadronPhysics3, Grant Agreement n.283286) under the Seventh 
Framework
Programme of the European Community and by the COPIN-IN2P3 Agreement.


\begin{appendix}
\section{Appendices}

\subsection{Distribution amplitudes in the LCCF parametrization}
\label{subsubsec:DA_LCCF}

The seven chiral-even\footnote{The chiral-odd twist-2 DA for the transversely polarized $\rho$ meson does
not contribute to the process considered at the accuracy discussed here. This is also true in the approach
based on collinear factorization of generalized parton distributions \cite{Diehl:1998pd, Collins:1999un}.}
$\rho$-meson DAs up to twist 3 are defined through\footnote{In the approximation where the mass of the quarks is neglected with respect to the mass of the $\rho$ meson.} the following matrix elements of nonlocal 
light-cone operators \cite{Anikin:2009bf}, for two-parton
\beqa
\label{defDA_V}
&& \left\langle \rho(p_{\rho}) \left|\bar{\psi}(z)\gamma_{\mu}\psi(0)\right|0\right\rangle  = m_{\rho} f_{\rho}  \int^{1}_{0} dy \, e^{i yp.z}  [\varphi_1(y;\mu_F^2) (e^*.n)p_{\mu}+\varphi_3(y;\mu_F^2)e_{T\mu}^*]\,, \\
\label{defDA_Vder}
&& \left\langle \rho(p_{\rho}) \left|\bar{\psi}(z)\gamma_5\gamma_{\mu}\psi(0)\right|0\right\rangle  =i \, m_{\rho} f_{\rho} \, R^*_{\perp\mu} \int^{1}_{0} dy \, e^{i yp.z}  \varphi_A(y;\mu_F^2)\,,\\
\label{defDA_A}
&& \left\langle \rho(p_{\rho}) \right|\bar{\psi}(z)\gamma_{\mu}i
\stackrel{\longleftrightarrow}
{\partial^T_{\alpha}}
\psi(0)\left|0\right\rangle  =m_{\rho} f_{\rho} \, p_{\mu}e^{*}_{T\alpha} \int^{1}_{0} dy \,  e^{i yp.z}  \varphi_1^T(y;\mu_F^2)\,,\\
\label{defDA_Ader}
&& \left\langle \rho(p_{\rho}) \right|\bar{\psi}(z)\gamma_5\gamma_{\mu}i\stackrel{\longleftrightarrow}
{\partial^T_{\alpha}}
\psi(0)\left|0\right\rangle  = i \, m_{\rho} f_{\rho} \, p_{\mu}\, R^*_{\perp \alpha}\int^{1}_{0} dy \, e^{i yp.z}  \varphi_A^T(y;\mu_F^2)
\,,
 \eqa
and  three-parton correlators 
 \beqa
\label{defDA_B}
&&\left\langle \rho(p_{\rho}) \left|\bar{\psi}(z_1)\gamma_{\mu} g A^T_{\alpha}(z_2)\psi(0)\right|0\right\rangle  = m_{\rho} f^{V}_{3\rho}(\mu_F^2) \,
p_{\mu} \, e_{T\alpha}^*\\
&&\hspace{2cm}\times \int^{1}_{0}dy_2\int^{y_2}_{0} d y_1 e^{i y_1p.z_1+i(y_2-y_1)p.z_2}  B(y_1,y_2;\mu_F^2) \nn\\
\label{defDA_D}
&&\left\langle \rho(p_{\rho}) \left|\bar{\psi}(z_1)\gamma_5\gamma_{\mu} g A^T_{\alpha}(z_2)\psi(0)\right|0\right\rangle = 
i \, m_{\rho} f^{A}_{3\rho}(\mu_F^2) \,
p_{\mu}\,R^*_{\perp \alpha}\\
&&\hspace{2cm}\times \int^{1}_{0}\!\!\!\!dy_2 \! \int^{y_2}_{0} \!\!\!\!d y_1 \, e^{i y_1p.z_1+i(y_2-y_1)p.z_2}  D(y_1,y_2;\mu_F^2) \,,\,\,\nn
\eqa
where we used the standard notation
$\stackrel{\longleftrightarrow}
{\partial_{\rho}}=\frac{1}{2}(\stackrel{\longrightarrow}
{\partial_{\rho}}-\stackrel{\longleftarrow}{\partial_{\rho}})\,.$

DAs are linked by linear differential relations derived from equations of motion and $n-$independency condition \cite{Anikin:2009hk,Anikin:2009bf}. The solutions for $\varphi_P(y)\equiv\{\varphi_3, \varphi_A, \varphi_1^T, \varphi_A^T\}$ are the sum of the solutions in the so-called WW approximation and  of genuine solutions,
\begin{equation}
 \varphi_P(y)=\varphi^{WW}_P(y)+\varphi^{gen}_P(y)\,.
 \end{equation}
The WW approximation consists in neglecting the contribution from three-parton operators, thus taking $B(y_1,y_2;\mu_F^2)=D(y_1,y_2;\mu_F^2)=0$. Then, $\varphi_P^{WW}(y)$ become functions of $\varphi_1(y)$ only, and their explicit expressions are given by
\beqa
\label{phi3_WW}
 \varphi^{WW}_{3}(y;\mu_F^2)&=&\frac{1}{2} \left[\int^{y}_{0}du \frac{\varphi_1(u;\mu_F^2)}{\bar{u}}+\int^{1}_{y} du \frac{\varphi_1(u;\mu_F^2)}{u}\right]\,,\\
\label{phiA_WW}
 \varphi^{WW}_{A}(y;\mu_F^2)&=&\frac{1}{2}\left[\int^{y}_{0}du \frac{\varphi_1(u;\mu_F^2)}{\bar{u}}-\int^{1}_{y} du \frac{\varphi_1(u;\mu_F^2)}{u}\right]\,,\\
\label{phiAT_WW}
\varphi^{T\,WW}_{A}(y;\mu_F^2)&=&-\frac{1}{2}\left[\bar{y} \int^{y}_{0}du \frac{\varphi_1(u;\mu_F^2)}{\bar{u}}+y\int^{1}_{y} du \frac{\varphi_1(u;\mu_F^2)}{u}\right]\,,\\
\label{phi1T_WW}
\varphi^{T\,WW}_{1}(y;\mu_F^2)&=&\frac{1}{2}\left[-\bar{y} \int^{y}_{0}du \frac{\varphi_1(u;\mu_F^2)}{\bar{u}}+y \int^{1}_{y} du \frac{\varphi_1(u;\mu_F^2)}{u}\right]\,.
\eqa
Genuine solutions only depend on $\{B(y_1,y_2;\mu_F^2), D(y_1,y_2;\mu_F^2)\}$ or equivalently on the combinations 
$\{S(y_1,y_2;\mu_F^2), M(y_1,y_2;\mu_F^2)\}$ defined by eq.~(\ref{defSM}), namely
\beqa
\label{exp_phi3gen}
\varphi^{gen}_3(y;\mu_F^2)&=&\frac{1}{2}\left[\int^{1}_{\bar{y}}d u\,\frac{A(u;\mu_F^2)}{u}+\int^{1}_{y}d u \, \frac{A(u;\mu_F^2)}{u} \right]
\\
\label{exp_phiAgen}
 \varphi^{gen}_A(y;\mu_F^2)&=&\frac{1}{2}\left[\int^{1}_{\bar{y}}d u \,\frac{A(u;\mu_F^2)}{u}-\int^{1}_{y}d u \,\frac{A(u;\mu_F^2)}{u} \right]\,,
 \eqa
where $A(u;\mu_F^2)$ has the compact form
\begin{equation}
A(u;\mu_F^2)=\int^{u}_{0}d y_2\left[\frac{1}{y_2-u}-\partial_u\right]M(y_2,u;\mu_F^2)+\int^{1}_{u}dy_2 \frac{1}{y_2-u}M(u,y_2;\mu_F^2)
\end{equation}
and  it obeys the conditions
\begin{equation}
\int^{1}_{0}du \, A(u;\mu_F^2)=0 \quad {\rm and} \quad \int^{1}_{0}du \,\bar{u} \, A(u;\mu_F^2)=0\,,
\end{equation}
coming, respectively, from the constraints
\beq
\label{int_phi3_phiA}
\int^1_0 \varphi_3^{gen}(y; \, \mu_F^2)  \, dy =0 \quad {\rm and} \quad  \int^1_0 (y-\bar{y}) \, \varphi_A^{gen}(y; \, \mu_F^2)  \, dy =0\,.
\eq
Equations (\ref{exp_phi3gen}) and (\ref{exp_phiAgen}) determine the expressions of $\varphi^{gen}_{1T}(y;\mu_F^2)$ and $\varphi^{gen}_{AT}(y;\mu_F^2)$ as
\beqa
\label{exp_phi1Tgen}
\varphi^{T\,gen}_{1}(y;\mu_F^2)&=&\int^{y}_{0}du\, \varphi^{gen}_3(u;\mu_F^2)\\
&&-\frac{1}{2} \int^{y}_{0}dy_1\int^{1}_{y}dy_2 \frac{S(y_1,y_2;\mu_F^2)+M(y_1,y_2;\mu_F^2)}{y_2-y_1}\,,\nn
\\
\label{exp_phiATgen}
\varphi^{T\,gen}_{A}(y;\mu_F^2)&=&\int^{y}_{0}du\,\varphi^{gen}_A(u;\mu_F^2)\\
&&-\frac{1}{2} \int^{y}_{0}dy_1\int^{1}_{y}dy_2 \frac{S(y_1,y_2;\mu_F^2)-M(y_1,y_2;\mu_F^2)}{y_2-y_1}\,.\nn
\eqa
The correspondence between our set of DAs and the one defined in ref.~\cite{Ball:1998sk}
is achieved through the following dictionary derived in ref.~\cite{Anikin:2009bf}. It reads,
for the two-parton vector DAs,
\begin{eqnarray}
\label{relBBvector}
&&\varphi_1(y)=
\phi_{\parallel}(y) ,
\quad
\varphi_3(y)=
 g_\perp^{(v)}(y) \,,
\end{eqnarray}
and for the axial DA,
\begin{eqnarray}
\label{relBBaxial}
&&\varphi_A(y) =
-\frac{1}{4} \, \frac{\partial g_\perp^{(a)}(y)}{\partial y}\,.
\end{eqnarray}
For the three-parton DAs, the identification is
\begin{eqnarray}
\label{DictB_D}
 B(y_1,\,y_2)=-\frac{V(y_1, \, 1-y_2)}{y_2-y_1} \quad {\rm and } \quad
D(y_1,\,y_2)=-\frac{A(y_1, \, 1-y_2)}{y_2-y_1}\,.
\end{eqnarray}
Explicit forms for $\varphi_1$, $B$, and $D$ are obtained with the help of the results of ref.~\cite{Ball:1998sk} obtained within the QCD sum rules approach.
The first terms of the expansion in the momentum fractions of the three independent DAs thus have the form
 \beqa
\label{asymp_phi1}
 \varphi_1(y,\mu_F^2)&=&6\, y\bar{y} \left[1+a_2^{\|}(\mu_F^2)\frac{3}{2} (5 (y-\bar{y})^2-1)\right]\,,\\
\label{asymp_B}
 B(y_1,y_2;\mu_F^2)&=&-5040 \, y_1 \bar{y}_2 (y_1-\bar{y}_2) (y_2-y_1)\,,\\
D(y_1,y_2;\mu_F^2)&=&-360 \, y_1\bar{y}_2(y_2-y_1) \left[1+\frac{\omega^{A}_{\{1,0\}}(\mu_F^2)}{2}(7(y_2-y_1)-3) \right]\,.
\label{asymp_D}
\eqa
The dependences on the renormalization scale $\mu_F$ of the coupling constants $a_2^{\|}$, $\omega^{A}_{\{1,0\}}$, $\zeta^{A}_{3}$, and $\zeta^{V}_{3}$ are given in ref.~\cite{Ball:1998sk}. 
In appendix \ref{subsec:DA_evolution} we present both the evolution equations and the 
values of these constants at $\mu_F^2=1$~GeV$^2$ used in our analysis,  as well as the dependence on $\mu_F$ of the DAs.

\subsection{Evolutions of DAs and coupling constants with the renormalization scale}
\label{subsec:DA_evolution}
 
The parameters entering the DAs at $\mu_0^2=1\;$GeV$^2$ are updated\footnote{We use the notations of ref.~\cite{Ball:1998sk} for the parameters, they are related to the updated parameters of ref.~\cite{Ball:2007zt} by the following relations, $\zeta_{3 }^A=\zeta^{\|}_{3}$, $\zeta_{3}^{V}=\omega_3^{\|}/14$ and $\zeta_{3}^{\|}\omega_{\{1,0\}}^{A}=\tilde{\omega}^{\|}_{3}$.} in ref.~\cite{Ball:2007zt} and their evolution equations are given in ref.~\cite{Ball:1998sk}, we recall in table~\ref{table:DAs} their values for the $\rho$ meson.
\begin{table}[htbp]
	\centering
 \begin{tabular}{|l|c|r|}
\hline
 $\alpha_s$ & 0.52\\ \hline
& \\
$\omega^{A}_{\{1,0\}}$ & -3.0\\ \hline
&\\
$\omega^{V}_{[0,1]}$ & 28/3\\ \hline
&\\
$a^{\|}_{2}$ & 0.15 
\\ \hline
&\\
$\zeta^{A}_{3}$ & 0.030\\ \hline
&\\
$\zeta^{V}_{3}$ & 0.011\\ \hline
\end{tabular}
\caption{Coupling constants and Gegenbauer coefficients entering the $\rho$ meson DAs, at the scale $\mu_0=1$ GeV updated in ref.~\cite{Ball:2007zt}. Note that in ref.~\cite{Ball:1998sk} the normalization are such that $f^{V,A}_{3\rho\, \mbox{\cite{Ball:1998sk}}}=m_\rho \, f^{V,A}_{3\rho\, [{\rm here}]}$.}
\label{table:DAs}
\end{table}

For $a_2^{\|}$, the evolution equation is
\begin{equation}
\label{a2evolv}
a_2^{\|}(\mu^2)= a_{2}^{\|}(\mu_0^2) \, L(\mu^2)^{\gamma_2/b_0}
\end{equation}
with
\begin{equation}
\label{defL}
L(\mu^2) = \frac{\alpha_s(\mu^2)}{\alpha_{s}(\mu_0^2)} = \frac{1}{1+\frac{b_0}{\pi} \alpha_{s}(\mu_0^2) \ln(\mu^2/\mu_0^2)}
\end{equation}
where $b_0=(11 N_c-2 N_f)/3\,,$
$\gamma_n=4 C_F\left (\psi(n+2)+\gamma_E-\frac{3}{4}-\frac{1}{2 (n+1)(n+2)}\right)$
and
$\psi(n) = -\gamma_E + \sum^{n+1}_{k=1} 1/k\,.$
For the $f^{A}_{3\rho}$ coupling constant, the evolution is given by
\begin{equation}
\label{fAevolv}
 f^{A}_{3\rho}(\mu^2)=f^{A}_{3\rho}(\mu_0^2)\, L(\mu^2)^{\Gamma^{-}_2/b_0}
 \end{equation}
with $\Gamma_2^- = -\frac{C_F}{3}+3 \, C_g$ ($C_g = N_c$).
The couplings
$f^{V}_{3\rho}$ and $\omega^{A}_{ \{ 0,1\} }(\mu^2) f^{A}_{3\rho}(\mu^2)$ enter a matrix evolution equation \cite{Ball:1998sk}. Defining
\beqa
\label{defV}
 V(\mu^2)=\left(
 \begin{array}{clrr}
       \omega^{V}_{[0,1]} f^{V}_{3\rho}(\mu^2) -  \omega^{A}_{ \{ 0,1\} }(\mu^2) f^{A}_{3\rho}(\mu^2)\\
\\
       \omega^{V}_{[0,1]} f^{V}_{3\rho}(\mu^2) + \omega^{A}_{ \{ 0,1\} }(\mu^2) f^{A}_{3\rho}(\mu^2)
\end{array}
\right)\,,
\eqa
it reads
\begin{equation}
\label{V_ev}
V(\mu^2)=L(\mu^2)^{\Gamma^{+}_3/b_0} V(1)\,,
\end{equation}
with $\Gamma^{+}_3$ given by
\beq
\label{defGamma3}
 \Gamma^{+}_3=\left(
 \begin{array}{clrr}
       \frac{8}{3}C_F+\frac{7}{3}C_g & \ \frac{2}{3}C_F-\frac{2}{3}C_g \\
\\
       \frac{5}{3}C_F-\frac{4}{3}C_g & \ \frac{1}{6}C_F+4 C_g
\end{array}
\right)\,.
\eq
Hence we get the dependence of $f^{V}_{3\rho}$ and $\omega^{A}_{ \{ 0,1\} }$ by diagonalizing the system. 
 The dependence of the DAs on the renormalization scale is shown in ref.~\cite{Anikin:2011sa}. The DAs exhibit a non-negligible effect of QCD evolution, in particular, for the genuine
twist-3 contributions. We recall that we chose the collinear factorization scale of production of the $\rho$ meson $\mu_F$ to be equal to the renormalization scale of the process $\mu$, thus the dependence of the coupling constant in $\mu_F$ is given by eqs.~(\ref{a2evolv}, \ref{fAevolv}, \ref{V_ev}).

\subsection{Dipole-proton scattering amplitude in the GS-Model}
\label{subsec:DipProtonGS}
In this appendix we calculate the Fourier transform of the proton impact factor. We denote by $r=\left|\rb \right|$ the transverse size of the dipole. We should thus compute
\beq
\sigma(r)= A \int d^2 \kb \frac{1}{(\kb^2)^2} \left(\frac{1}{M^2} - \frac{1}{M^2+\kb^2}\right) (1 - e^{i \kb \cdot \rb}) (1 - e^{-i \kb \cdot \rb}) \,.
\label{def-fourier-impact}
\eq
Using the notation $k=|\kb|\,,$ the angular integration leads to
\beq
\sigma(r)=4 \pi \, A \int d k \frac{1}{k^3} \left(\frac{1}{M^2} - \frac{1}{M^2+\kb^2}\right) (1 - J_0(k \, r)) \,.
\label{fourier-impact1}
\eq
Relying on the identity
\beq
\frac{1}{k^3} \left(\frac{1}{M^2} - \frac{1}{M^2+k^2}\right)=\frac{1}{M^4} \left(\frac{1}{k} - \frac{k}{M^2+k^2}\right)\,,
\label{identity-apart}
\eq
we rewrite $\sigma(r)$ as
\beq
\sigma(r)= \frac{4 \pi A}{M^4} I
\label{def-I-sigma}
\eq
with
\beqa
I&=&\int^\infty_0 \frac{d k }k  - \int^\infty_0   \frac{d k }k J_0(k \, r) - \int^\infty_0 \frac{k}{M^2+k^2} dk + \int^\infty_0 \frac{k}{M^2+k^2}  J_0(k \, r) \, dk \nonumber \\
&=&  I_1 - I_2 - I_3 + I_4\,.
\label{def-I_i}
\eqa
The integral $I_4$ is UV and IR finite and reads \cite{Gradshteyn}
\beq
I_4 = K_0(M  \, r)\,.
\label{I4}
\eq
The integrals $I_1$ and $I_2$ are both IR divergent and are regularized through  dimensional regularization, while the UV divergencies of $I_1$ and $I_3$ are regularized through a cut-off $\Lambda\,.$
We thus write
\beq
I_1 = \int^\Lambda_0   \frac{d k }{k^{1-\epsilon}} \sim \frac{1}\epsilon + \ln \Lambda\,. 
\label{I1}
\eq
Using the relation \cite{Gradshteyn}
\beq
\int^\infty_0 x^\mu \, J_\nu (a \, x) \, dx = 2^\mu a^{-\mu-1} \frac{\Gamma\left(  \frac{1}2 + \frac{1}2 \nu + \frac{1}2 \mu\right)}{\Gamma\left(  \frac{1}2 + \frac{1}2 \nu - \frac{1}2 \mu\right)}
\label{Grad-2}
\eq
we obtain
\beq
I_2 = \int^\infty_0   \frac{d k }{k^{1-\epsilon}} J_0(k \, r) = 2^{-1+\epsilon}\, r^{-\epsilon} \frac{\Gamma\left( \frac{\epsilon}2\right)}{\Gamma\left( 1-\frac{\epsilon}2\right)} \sim \frac{1}\epsilon - \ln r -\gamma + \ln 2\,. 
\label{I2}
\eq

Finally, 

\beq
I_3 = \int^\Lambda_0   \frac{k}{M^2+k^2} d k \sim \ln \Lambda - \ln M\,. 
\label{I3}
\eq

Thus, combining eqs.~(\ref{I1}, \ref{I2}, \ref{I3}, \ref{I4}) we get

\beq
I=\left(\gamma + \ln \frac{M \, r}2 + K_0(M  \, r)\right)\,,
\label{I-final}
\eq
and thus
\beq
\sigma(r)=\frac{4 \pi \, A}{M^4}  \left(\gamma + \ln \frac{M \, r}2 + K_0(M  \, r)\right)\,.
\label{sigma-r-final}
\eq

\subsection{Results using the GBW and AAMQSb models}
\label{subsec:GBW-AAb}

We present some of the predictions obtained by using the GBW or the AAMQSb model for the dipole cross-section. As expected the results are not so far from the ones obtained with the AAMQSa model. In figure~\ref{FigAAb1} and \ref{FigGBW}
 are respectively shown for the AAMQSb and the GBW model, the polarized cross-sections $\sigma_T$ and $\sigma_T$. The spin density matrix element $r^{04}_{00}$ predictions using these dipole models are shown in figure~\ref{Figr04b}, 
  for completeness we show also the prediction obtained with the GS model.

\def\taille{14}
\begin{figure}[htbp]
\hspace*{\fill}
\begin{tabular}{c}
\subfigure[$\sigma_T$  vs H1 data~\cite{Aaron:2009xp}.
]{\epsfxsize=\taille cm\epsfbox{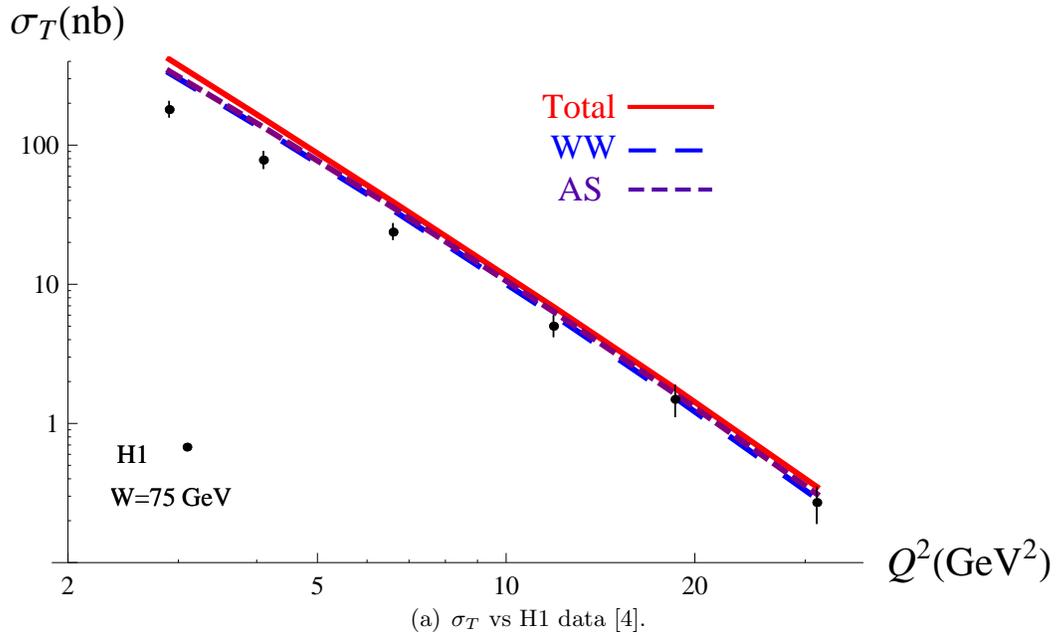}\label{sigTb}}\\
 \subfigure[$\sigma_L$  vs H1 data~\cite{Aaron:2009xp}.
]{\epsfxsize=\taille cm\epsfbox{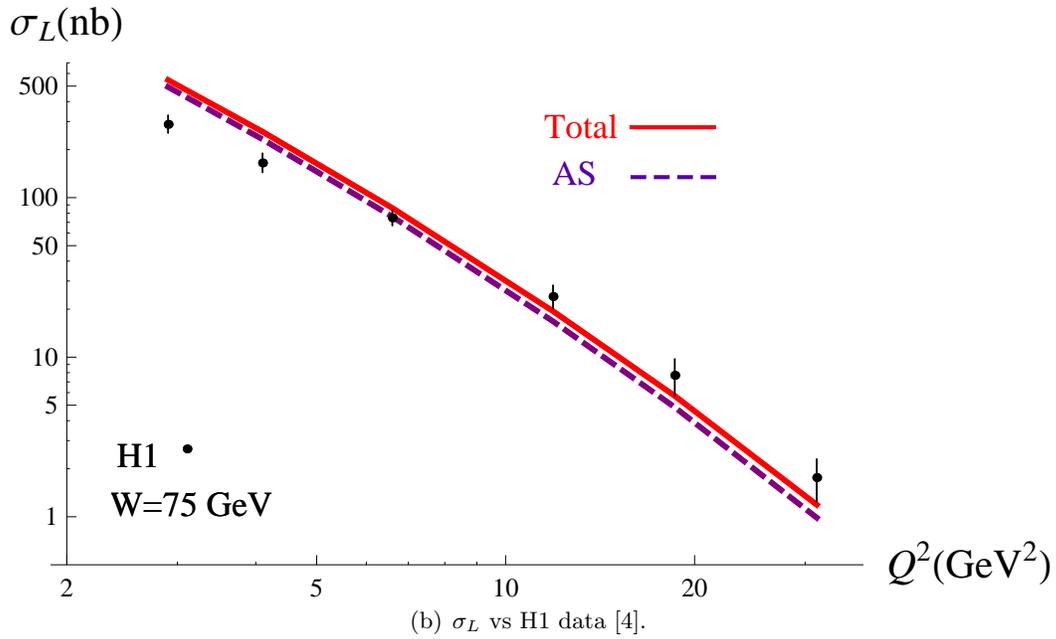}\label{sigLb}}
\end{tabular}
\hspace*{\fill}
\caption{AS (purple dashed lines), WW (blue long dashed lines) and Total (red solid lines) contributions to $\sigma_T$ and $\sigma_L$ vs $Q^2$, for $W=75\;$GeV, using the AAMQSb-model, compared to the data of H1\cite{Aaron:2009xp}.}
\label{FigAAb1}
\end{figure}

\newpage
\mbox{  }

\def\taille{14}
\begin{figure}[htbp]
\hspace*{\fill}
\begin{tabular}{c}
 \subfigure[$\sigma_T$  vs H1 data~\cite{Aaron:2009xp}.]{\epsfxsize=\taille cm\epsfbox{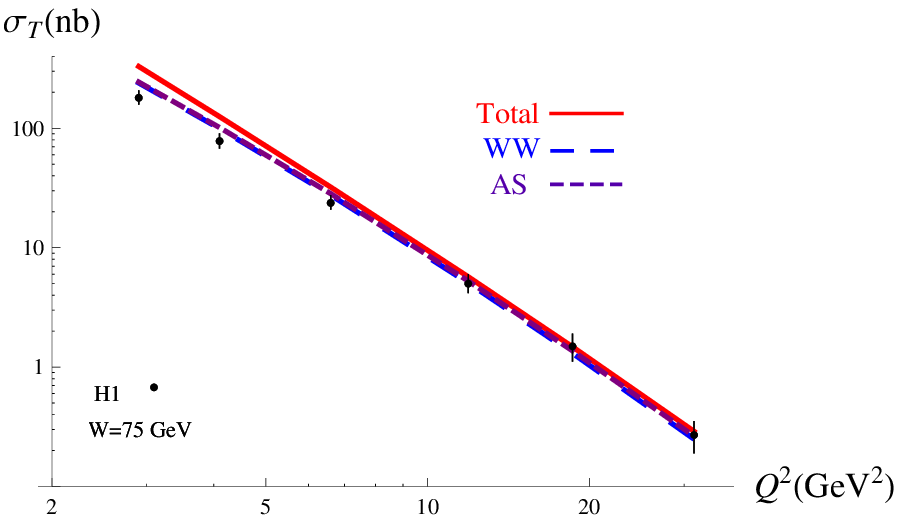}\label{sigTGBW}}\\
 \subfigure[$\sigma_L$  vs H1 data~\cite{Aaron:2009xp}.]{\epsfxsize=\taille cm\epsfbox{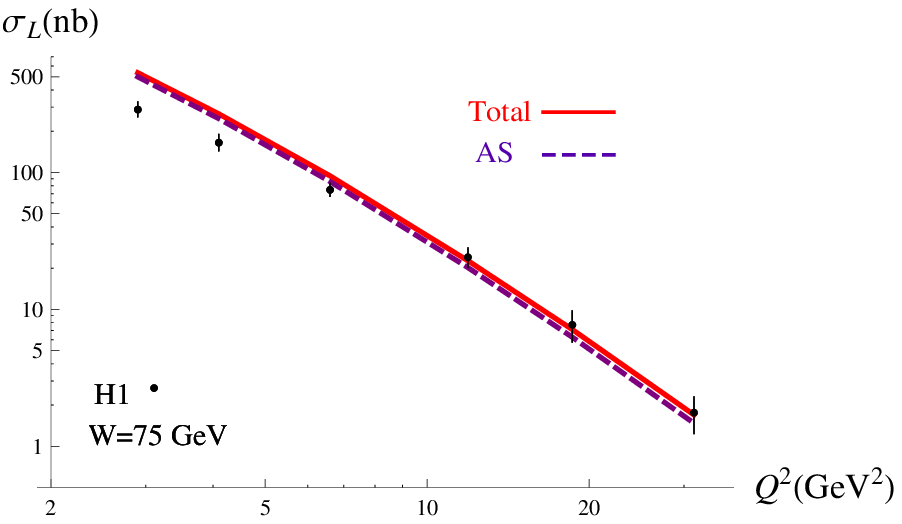}\label{sigLGBW}}
\end{tabular}
\hspace*{\fill}
\caption{AS (purple dashed lines), WW (blue long dashed lines) and Total (red solid lines) contributions to $\sigma_T$ and $\sigma_L$ vs $Q^2$, for $W=75\;$GeV, using the  GBW-model, compared to the data of H1\cite{Aaron:2009xp}.}
\label{FigGBW}
\end{figure} 

\newpage
\mbox{  }

\begin{figure}[htbp]
\hspace*{\fill}
\begin{tabular}[p]{cc}
\hspace{-.2cm}\subfigure[$r^{04}_{00}$ contributions using AAMQSb-model vs H1 data.]{\epsfxsize=7.3 cm\epsfbox{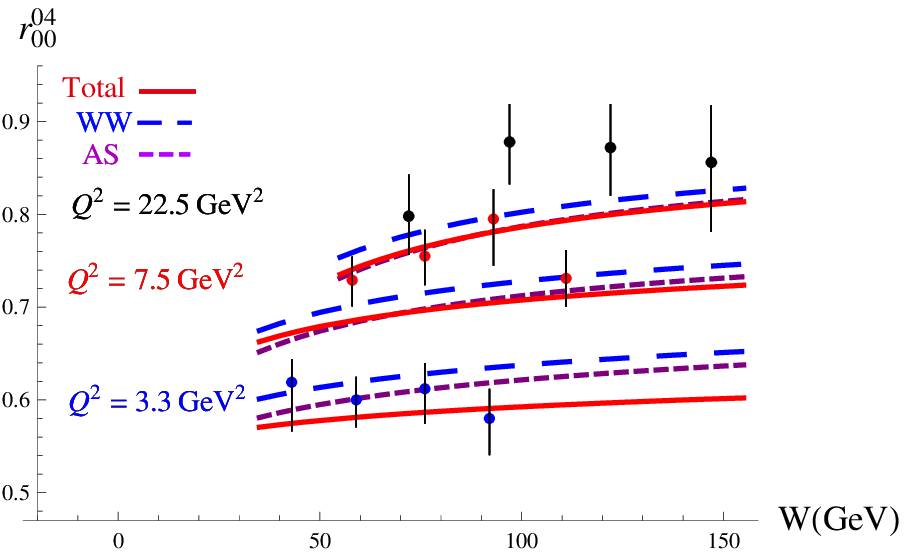}\label{rbh}}&
\hspace{-.2cm}
 \subfigure[$r^{04}_{00}$  contributions using AAMQSb-model vs ZEUS data.]{\epsfxsize=7.3 cm \epsfbox{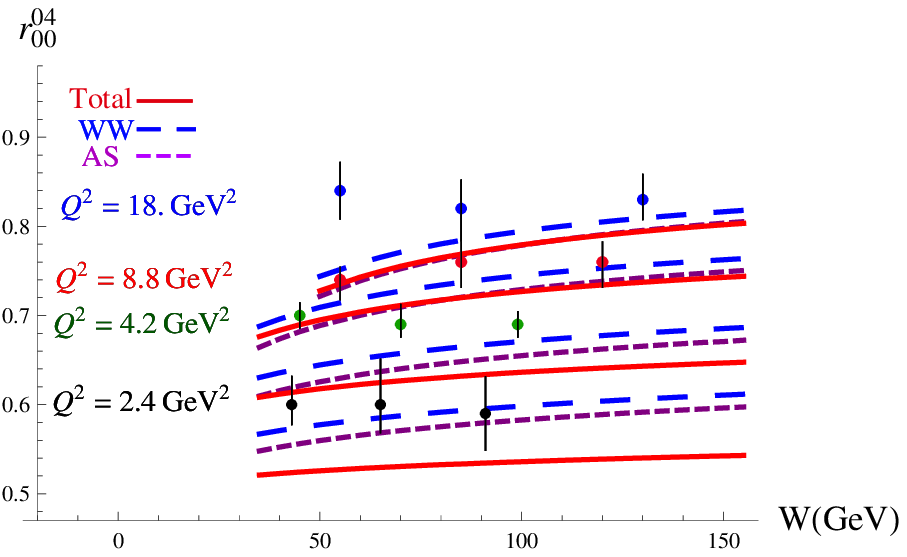}\label{rbz}}\\
\hspace{-.2cm} \subfigure[$r^{04}_{00}$ contributions using GBW-model vs H1 data.]{\epsfxsize=7.3 cm\epsfbox{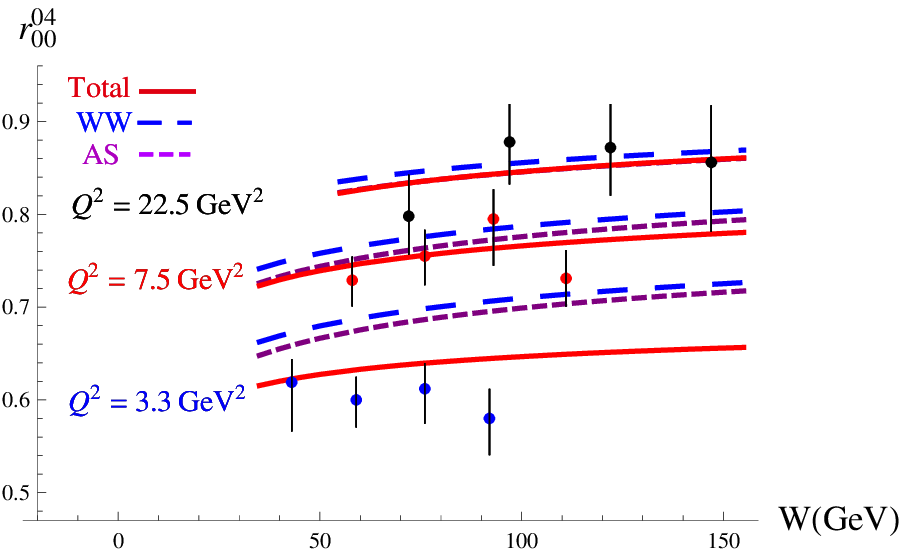}\label{rgbh}}&
\hspace{-.2cm} \subfigure[$r^{04}_{00}$ contributions using GBW-model vs ZEUS data.]{\epsfxsize=7.3 cm \epsfbox{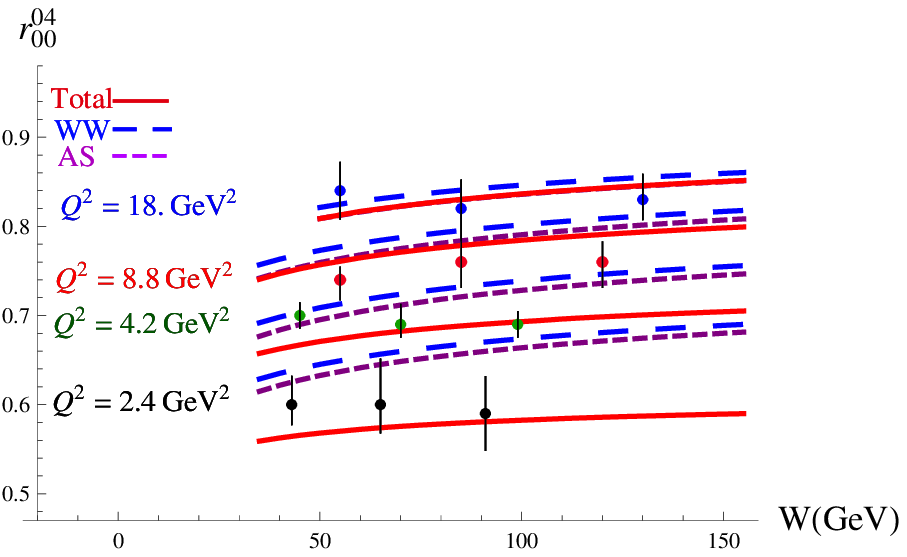}\label{rgbz}}\\
\hspace{-.2cm} \subfigure[$r^{04}_{00}$ contributions using GS-model vs H1 data.]{\epsfxsize=7.3 cm\epsfbox{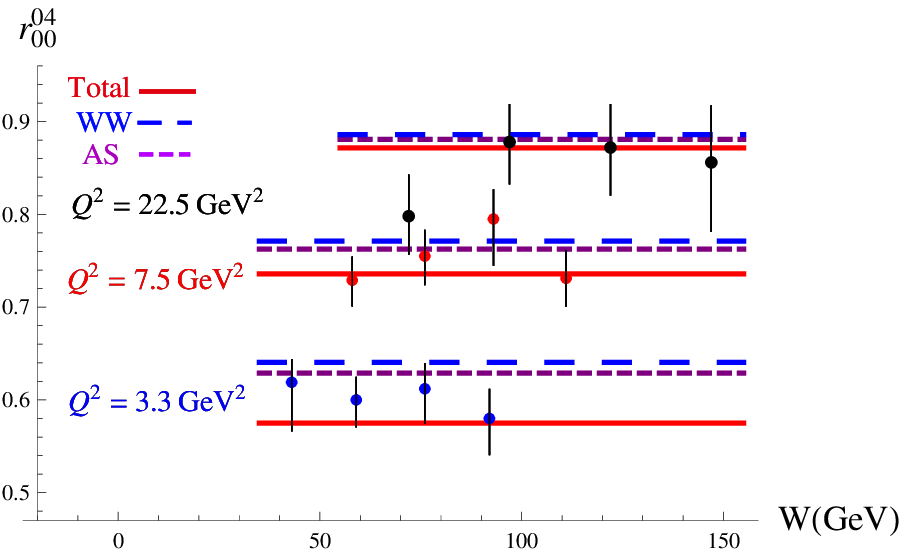}\label{rgsh}}&
 \hspace{-.2cm} \subfigure[$r^{04}_{00}$ contributions using GS-model vs ZEUS data.]{\epsfxsize=7.3 cm \epsfbox{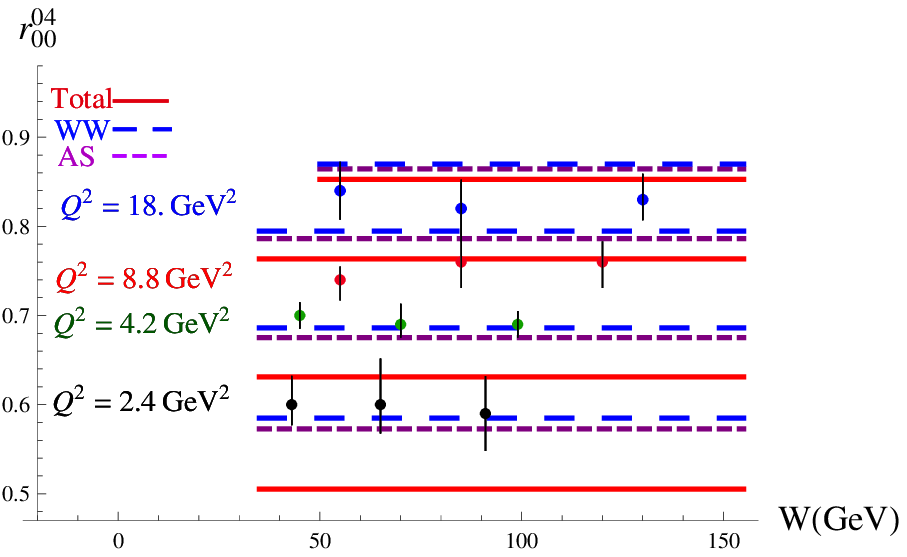}\label{rgsz}}
\end{tabular}
\hspace*{\fill}
\caption{Predictions for $r^{04}_{00}$ vs $W$ and $Q^2$ compared respectively with H1\cite{Aaron:2009xp} and ZEUS\cite{Chekanov:2007zr} data, the  AS (purple dashed lines), WW (blue long dashed lines), Total (red solid lines) contributions are shown separately using the AAMQSb-model, 
the GBW-model 
or the GS-model. 
}
\label{Figr04b}
\end{figure}

\end{appendix}

\newpage

\newpage

\providecommand{\href}[2]{#2}\begingroup\raggedright\endgroup

\end{document}